\documentclass[reprint,amsmath,amssymb,aip,showpacs,pra]{revtex4-1}
\usepackage{graphicx}
\usepackage{dcolumn}
\usepackage{bm}
\usepackage[usenames,dvipsnames]{color}
\usepackage[utf8]{inputenc}
\usepackage{hyperref}
\graphicspath{{./Files/PDFs/}{./Files/PNGs/}}

\begin{document}

\title{Viscoelastic effects on asymmetric two-dimensional vortex patterns in a strongly coupled dusty plasma}

\author{Akanksha Gupta}
\email{akgupt@iitk.ac.in}
\affiliation{Department of Physics, Indian Institute of Technology, Kanpur 208016, India}
\author{Rupak Mukherjee}
\email{rupak@ipr.res.in}
\author{Rajaraman Ganesh}
\email{ganesh@ipr.res.in}
\affiliation{Institute for Plasma Research, HBNI, Bhat, Gandhinagar - 382428, India}


\begin{abstract}
Strongly coupled dusty plasma medium is often described as a viscoelastic fluid that retains its memory. In
a flowing dusty plasma medium, vortices of different sizes appear when the flow does not remain laminar.
The vortices also merge to transfer energy between different scales. In the present work, we study the effect
of viscoelasticity and compressibility over a localized vortex structure and multiple rotational vortexes in a
strongly coupled viscoelastic dusty plasma medium. In case of single rotating vortex flow, a transverse wave
is generated from the localized vortex source and the evolution time of generated waves is found to be reduced
due to finite viscoelasticity and compressibility of the medium. It is found that the viscoelasticity
suppresses the dispersion of vorticity. In the presence of multiple vortices, we find, the
vortex mergers get highly affected in the presence of memory effect of the
fluid, and thus the dynamics of the medium gets completely altered compared to a non-viscoelastic fluid. For
a compressible fluid, viscoelasticity damps out the energy in the sonic waves generated in the medium. Thus
a highly viscoelastic and compressible fluid, in some cases, behaves similarly to an incompressible viscoelastic
fluid. The wave-front like rings propagate in elliptical orbits keeping the footprint of the earlier position of the
point-vortex. The rings collide with each other even within the patch vortex region forming regions of high
vorticity at the point of intersection and pass through each other.
\end{abstract}

\maketitle

\section{Introduction}
Dusty plasma is a widely studied medium not only for its huge applications in plasma physics\cite{morfill:2009, tsytovich:2007, fortov:2009, bonitz:2010, shukla:2015,morfillRMP,fusion,rao} but also for the study and understanding of colloids and non-newtonian fluids, like fluids with polymer concentration dynamics e.g, Hyaluronan/hyaluronic acid which comprises  long and unbranched polymer of various concentration  in synovial fluid etc \cite{pustejovska:2008}. Dusty plasma is a medium where the micron sized dust grains are immersed in a plasma and the dust grains form sheaths around each of them and collect a huge amount of negative charge from the plasma due to the higher mobility of electrons. Thus the charged dust particles interact between themselves very strongly through a screened coulomb or Yukawa potential. In a typical experiment, gravitational force acting on the dust particles gets balanced by the repulsive force felt by the dust grains from the cathode placed below the particles. Thus the dust grains get levitated in a horizontal plane forming a two dimensional fluid sheet. This offers a novel opportunity to study two dimensional turbulence in a strongly correlated charged fluid medium.  \\

The generalized hydrodynamic (GHD) model \cite{kaw:1998, diaw:2015, luo:2016, banerjee:2012, sanat:2012, akanksha:2014, akanksha:2017, akanksha:2018, sandeep:2017} has been proven to be one of the most successful models to describe the dynamics of such a medium. The model aptly takes care of the viscoelastic nature of the fluid through an additional time-derivative in the momentum equation. Recently experimental studies on flow related problems in a dusty plasma medium have attracted a lot of interests because of its inherent richness. \cite{saitou:2012, nakamura:1999, meyer:2013,  surabhi:2016} In case of a flowing dusty plasma, the medium acts like a viscoelastic fluid that possesses the features of memory of the medium. For a non-laminar flow of this dust fluid, eddies and vortices of different length scales get generated \cite{shukla:2003, hasegawa:2004, vranjevs:1999, wilms:2017} and merge during the flow. These mergers transfer energy through non-linear processes. For a memory based fluid, the energy transfer process gets significantly altered. The dynamics of such a viscoelastic fluid in the presence of shear has been already found to be quite different from that of a well-known Newtonian fluid. \cite{banerjee:2012, sanat:2012, akanksha:2014, akanksha:2017, akanksha:2018}\\

The evolution of coherent localized vortex structures is a very fundamental problem in the field of quasi-geotrophic and astrophysical phenomena \cite{krommes:2002}. In case of a turbulent flow driven at any intermediate scale, the merging of vortices producing larger vortex structure, fundamentally bifurcates the study from its three dimensional counterpart \cite{frisch:1995}. In two dimensions the inverse cascade of energy towards large scales, offers a completely different aspect of turbulence studied very well, analytically and of late numerically \cite{frisch:1995}. The aspect which is not that well-explored is - what happens when there is a memory associated with the fluid. The visco-elasticity alters the energy transfer mechanism from scale to scale and thereby posing several challenging questions in further understanding of such medium \cite{akanksha:2017}. For example, what happens to the rate of energy cascade in a fully developed turbulence in visco-elastic fluids is very hard to analyse. To answer this question, the basic nonlinear vortex merger studies need to be performed and well understood. Hence, in this work we primarily focuss on a simpler version of the problem asking the question that, how does different vortices time evolve in a visco-elastic fluid and what are the length scales associated with the assymmetric vortex mergers. We single out a case of such asymmetric vortex merger which is well studied in regular hydrodynamics \cite{rupak:2018} and study the time evolution of such processes delineating the differences from the hydrodynamic one.\\

  Experimental evidences of such two-dimensional fluid phenomena that occur
when intense, pointlike vortices are placed within a diffuse, circular vortex, can be found in earlier experimental works regarding the relaxation of turbulent magnetised electron column \cite{fine:1995, durkin:2000, schecter:1999, driscoll:1999}. Such vortex merger processes for a non-viscoelastic neutral fluid are extensively studied in past both at incompressible \cite{ganesh:2002,perlekar:2016} and compressible regimes \cite{rupak:2018}. In this report we study the evolution dynamics of localized vortex flow and  merging of  asymmetric vortices in a strongly coupled dusty plasma where non-newtonian nature is prominent. In present work, we study the effect of viscoelasticity and compressibility over the single and multiple rotational vortex  in a strongly coupled viscoelastic medium similar to dusty plasma. We found the propagtion of transverse waves for the evolution of a single Gaussian vortex in which viscoelasticity delays the generation of transverse shear waves. \\
  
  In the presence of multiple vortex flows, it was found that the central patch vortex becomes unstable and generates a ``V" state through the excitation of surface waves also called ``Kelvin waves". Then ``filamentation" process happen and extended ``fingers" are drawn out of the patch vortex\cite{rupak:2018}. At the non-linear phase, the filaments ``wave-break" and engulfs the vortices generating some quasistationary vortex-hole pattern. In this paper, first we reproduce the previously studied case for a charged fluid with an almost non-viscoelastic fluid ($\tau_m \rightarrow 0$ limit) . In case of a viscoelastic fluid, the whole merger process is found to get altered and thereby leading to completely new dynamics of the system. \\
  
The present paper has been divided in the following sections. In the Sec.~\ref{sec-goveq} the non-dimensionalized governing equation of strongly coupled dusty plasma has been dealt with. In the Sec.~\ref{sec-simulation}, simulation details and initial condition of single and multiple vortex flows have been explained. Sec.~\ref{sec-result} is for the discussion about the results obatined by the nonlinear evolution of single Gaussian, Rankine and multiple Rankine vortex patches. In the last Sec.~\ref{sec-summar}, we summarize our results and important findings of the work.


\section{Governing equations (Set of fluid equations)}

\label{sec-goveq}
Strongly coupled dusty plasmas can be often modelled by generalised hydrodynamics model\citep{kaw,akanksha:2018} within the range of fluid limit (for $1 < \Gamma < \Gamma_c$)$\:$\citep{vlad}, here $\Gamma_c$ is the liquid to solid phase transition point.
We consider only the dynamics of dust fluid medium and the background plasma medium is considered to respond in Boltzmann-like fashion. The set of non-dimensional nonlinear generalised hydrodynamics equations are as follows:

\begin{equation}
\frac{\partial n_{d}}{\partial t}+\nabla.(n_{d} \vec{U})=0 
\label{eq-nonden}
\end{equation}
\begin{equation}
\nabla^{2}\phi=n_{d}+\phi
\label{eq-poiss}
\end{equation}
\begin{equation}
\left\lbrace  1+\tau_{m}\frac{d}{dt} \right\rbrace  \left[ \frac{d\vec{U}}{dt}-\vec{\nabla}\phi +\frac{C_{s}^{2}\vec{\nabla} n_{d}}{n_{d}}\right] = \nu\nabla^{2}\vec{U}
\label{eq-nonmom}
\end{equation}

In the above  set of equations, $\vec{U}$, $\tau_m$, $n_{d}$, $C_{s}$ ($C_{s}^{2}=\mu_{d}\gamma_{d}k_{B}T_{d}/Z_{d}T_{i}$), and   $\nu$  are velocity of the dust fluid,  viscoelastic coefficient, dust density,  sound speed of the system, and kinematic shear viscosity respectively. In the present fluid model all equations are dimensionless and the normalization quantities are given in table.\ref{tab2}.
\begin{table}[h!]
\centering
 \begin{tabular}{||c c c ||} 
 \hline
 S.No & Quantity & Normalized quantity \\ [0.5ex] 
 \hline\hline
 1. & Distance (r) & r/$\lambda_{Dmix}$  \\ 
 2. & Time (t) & t$\omega_{pd}$  \\
 3. & Potential ($\Phi$) & $\phi e/k_{B}T_{i}$\\
 4. & Density ($n_{\alpha0}$, $\alpha=e,i,d$) & $n_{\alpha0}/Z_{d}n_{d0}$\\
 \hline
 \end{tabular}
 \caption {Table for normalized quantities.  where  length $\lambda_{Dmix}^{2}=\frac{\epsilon_{0}k_{B}T_{i}}{Z_{d}e^{2}n_{d0}}\simeq \lambda_{Di}^{2}$, dust plasma frequency $\omega_{pd}=(Q_{d}^{2}n_{d0}/\epsilon_{0}M_{d}$). $Q_{d}=Z_{d}e$ and $M_{d}$ are charge and mass of single dust particle.} \label{tab2} 
\end{table}\\
\noindent  Note that for single dust grain, $n_{d}=\delta{\vec{r}}$. Hence Eq.\ref{eq-poiss} yields Yukawa potential. Equation of state for dust fluid is $P=\mu_{d}\gamma_{d}n_{d}k_{B}T_{d}$, where $\mu_{d}$ and  $\gamma_{d}$ are compressibility and adiabatic  index ($\gamma_{d}=c_{p}/c_{v}$) respectively. 


\section{Simulation details }
\label{sec-simulation}
To perform the computational fluid dynamics study of grain medium, the above non-linear coupled set of fluid equations have been solved numerically  using pseudo spectral method through FFTW libraries\cite{FFTW3:2005}. For this purpose, a massively parallelized Advanced Generalised SPECTral Code (AG-Spect) has been developed and benchmarked against linear eigen value solver. Advanced Generalized SPECTral Code (AG-Spect) is an efficient tool to solve any set of coupled nonlinear partial differential equations.\cite{akanksha:2017, akanksha:2018}. To remove the aliasing error, there are several methods which one can use such as $\frac{3}{2}$-rule\citep{iov}, $\frac{2\sqrt{2}}{3}$-rule,\citep{orszag} $\frac{2}{3}$-rule (zero-padding method)\citep{ea}. In this code zero-padding method  has been used. The discretizations in space and time  are such that Courant-Friedrichs-Lewy (CFL) condition is well satisfied. It is important to note that, in this time-dependent numerical study, we have considered linearised Poisson's equation for simplicity. Nordsieck predictor-corrector method has been used for time-stepping.\\

\section{Results and discussions}
\label{sec-result}
In the present section, we study the single and multiple vortex evolution dynamics in the presence of compressibilty and viscoelasticity. For this purpose, we chose Gaussian vortex, single Rankine and multiple small Rankine vortex around a large Rankine vortex as initial flow. It is known that the initial density profile can significantly affect the dynamics of the fluid system. \cite{terakado:2014,bayly:1992} However, for the sake of simplicity, we consider a uniform density fluid ($n_d(x,y,0) = n_{d_0}$) only to start with. 
\subsection{Single rotatating vortex: Gaussian vortex}
\begin{figure*}

    \includegraphics[width=5cm, height=4cm]{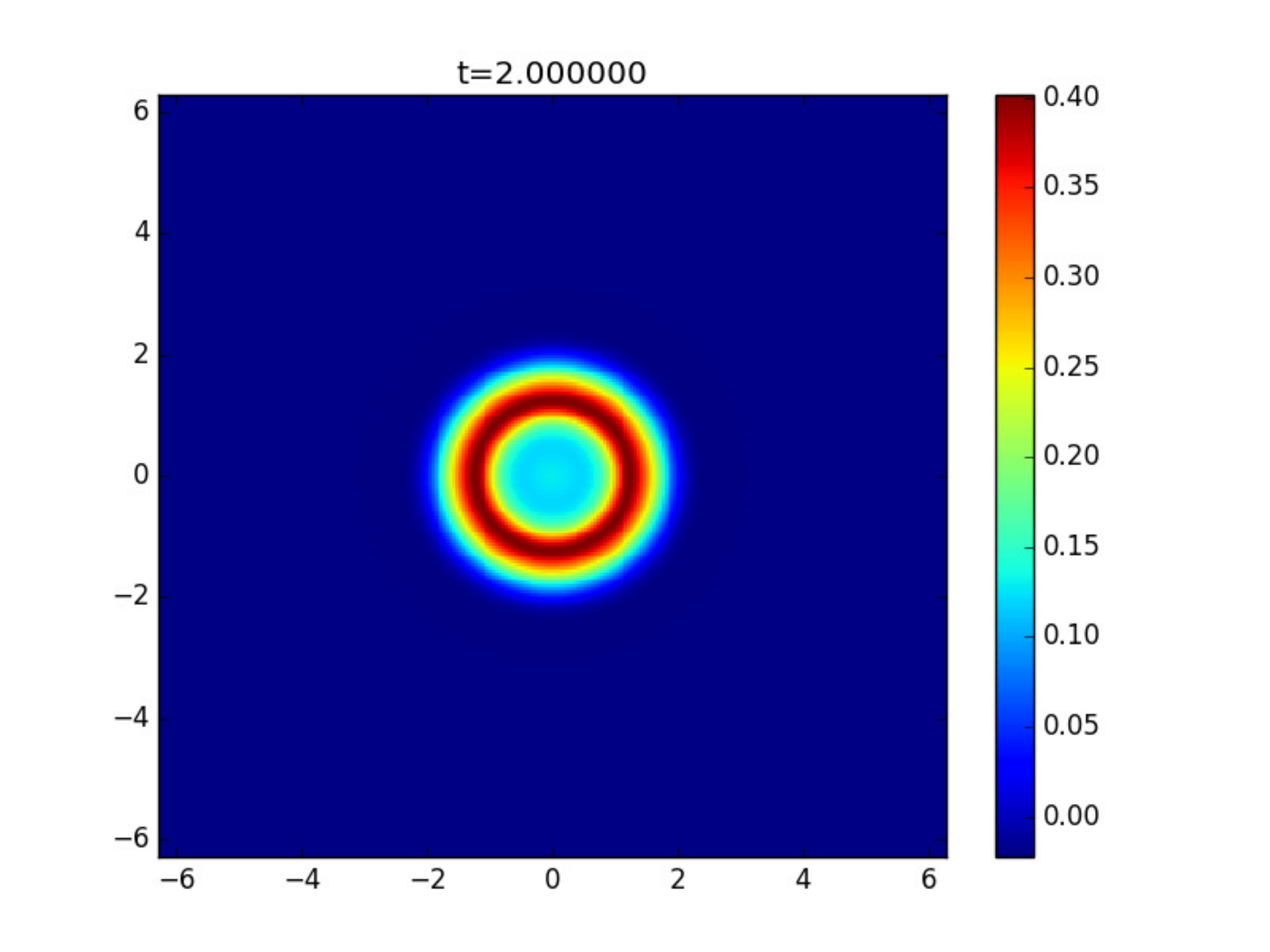} 
      \includegraphics[width=5cm, height=4cm]{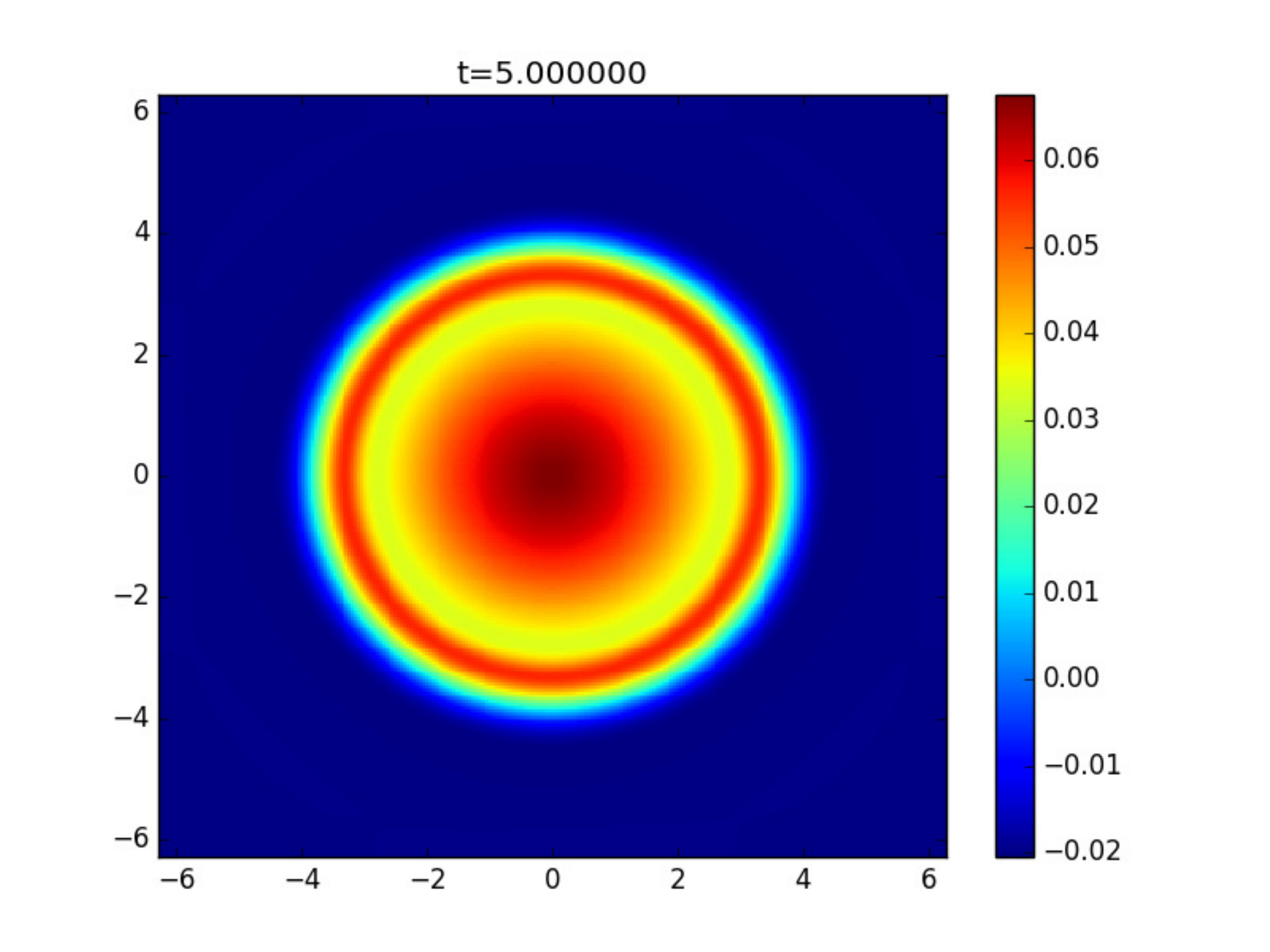}
       \includegraphics[width=5cm, height=4cm]{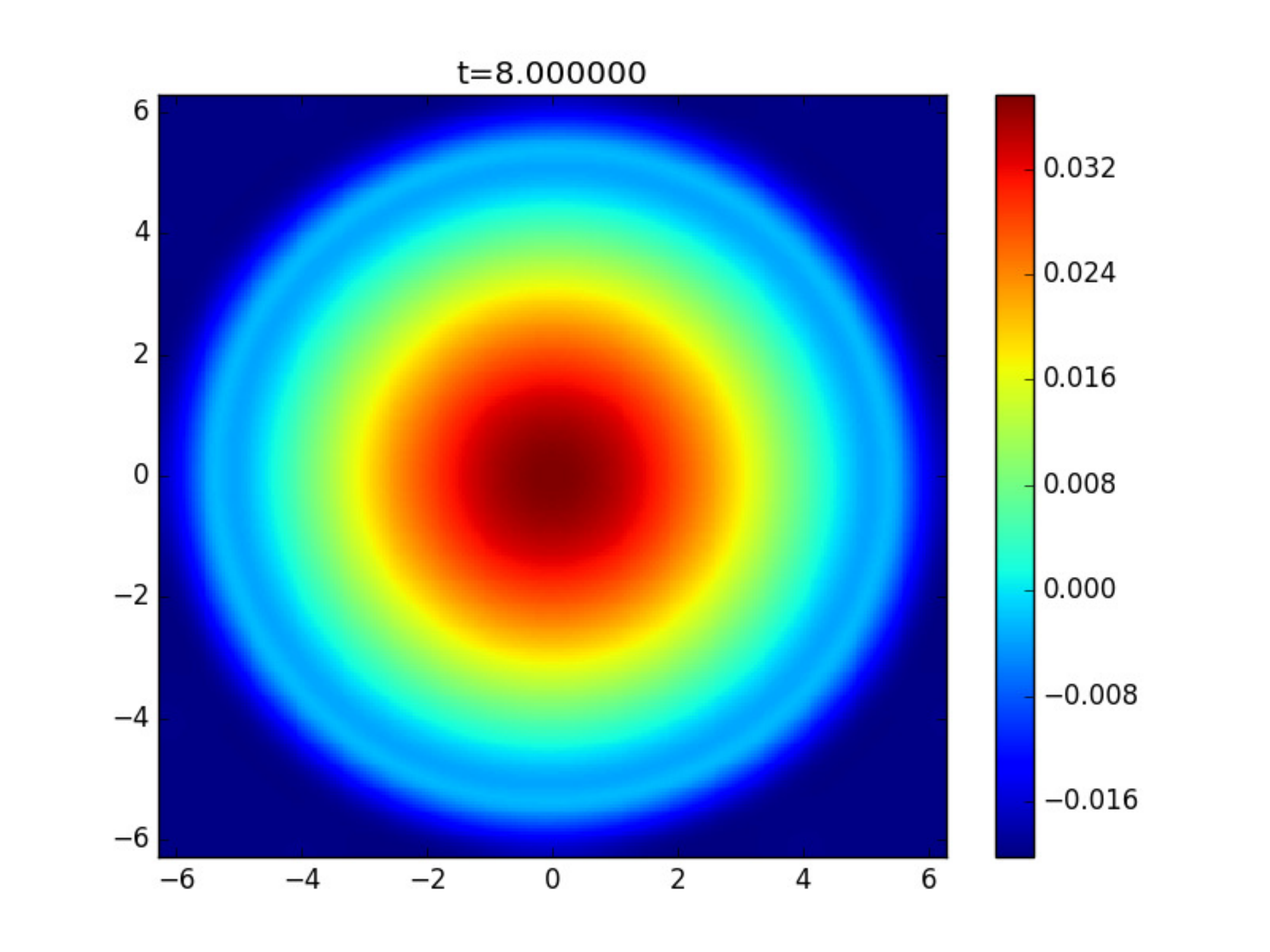}  \\    
     \includegraphics[width=5cm, height=4cm]{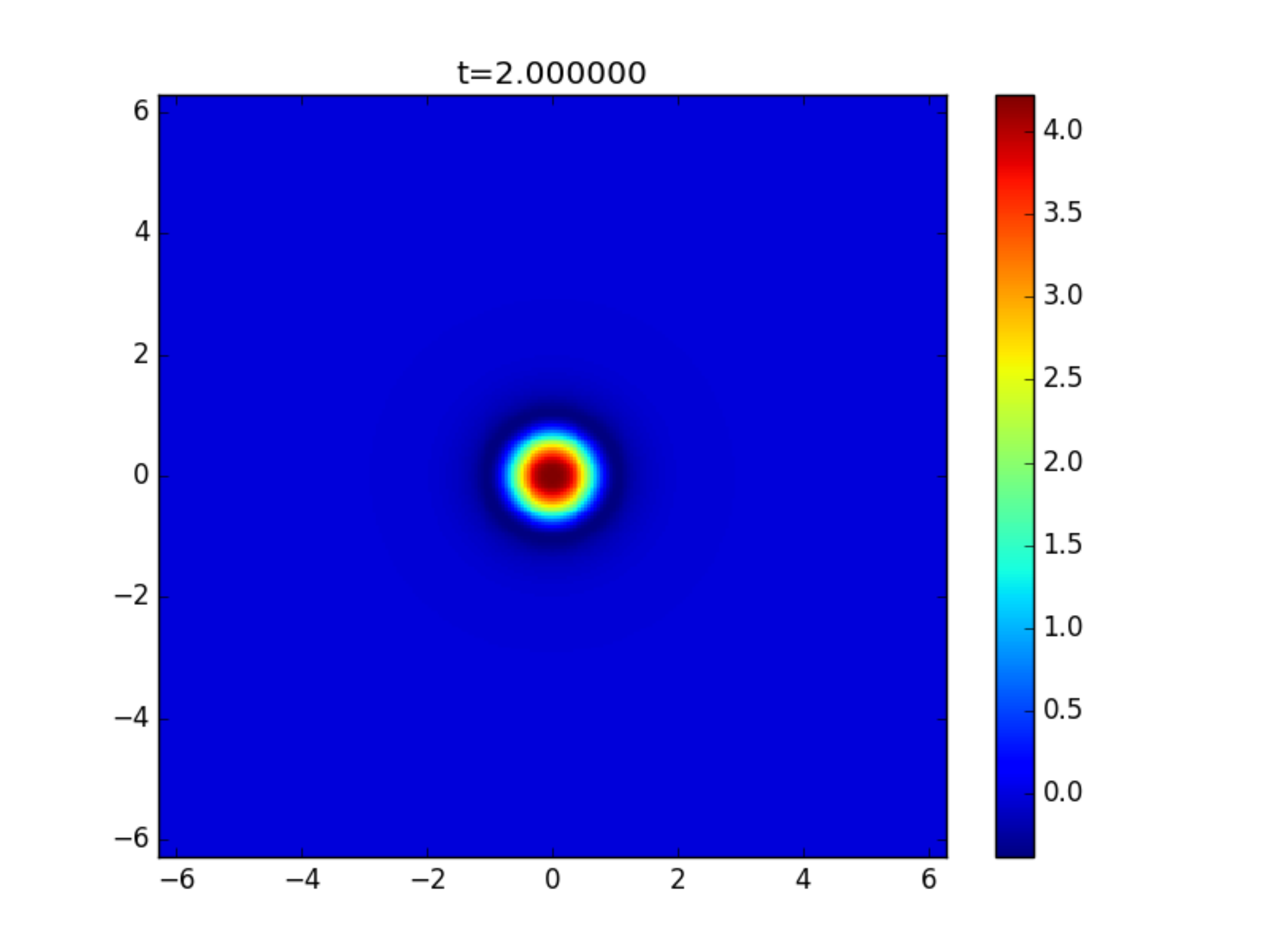} 
      \includegraphics[width=5cm, height=4cm]{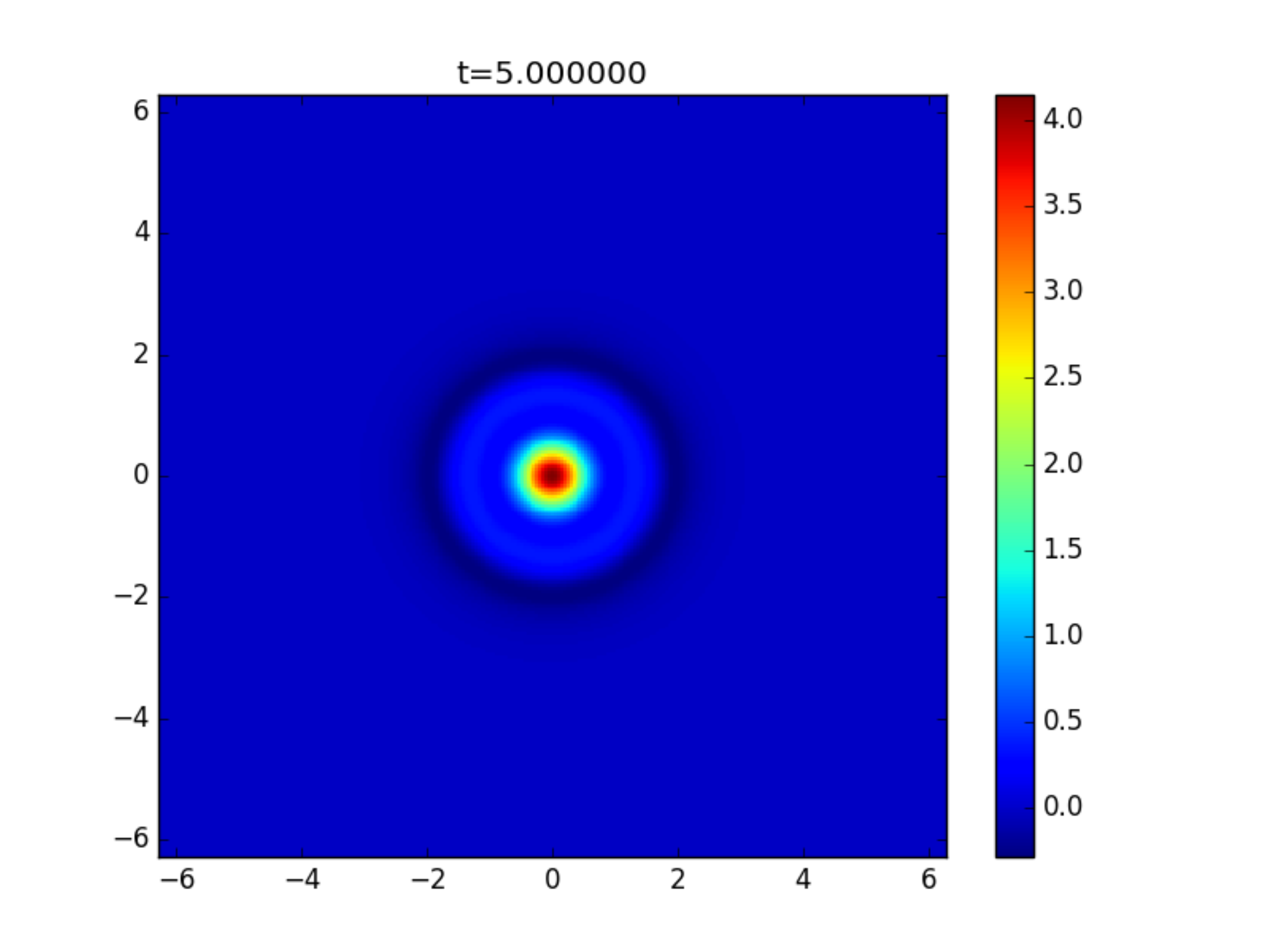}
       \includegraphics[width=5cm, height=4cm]{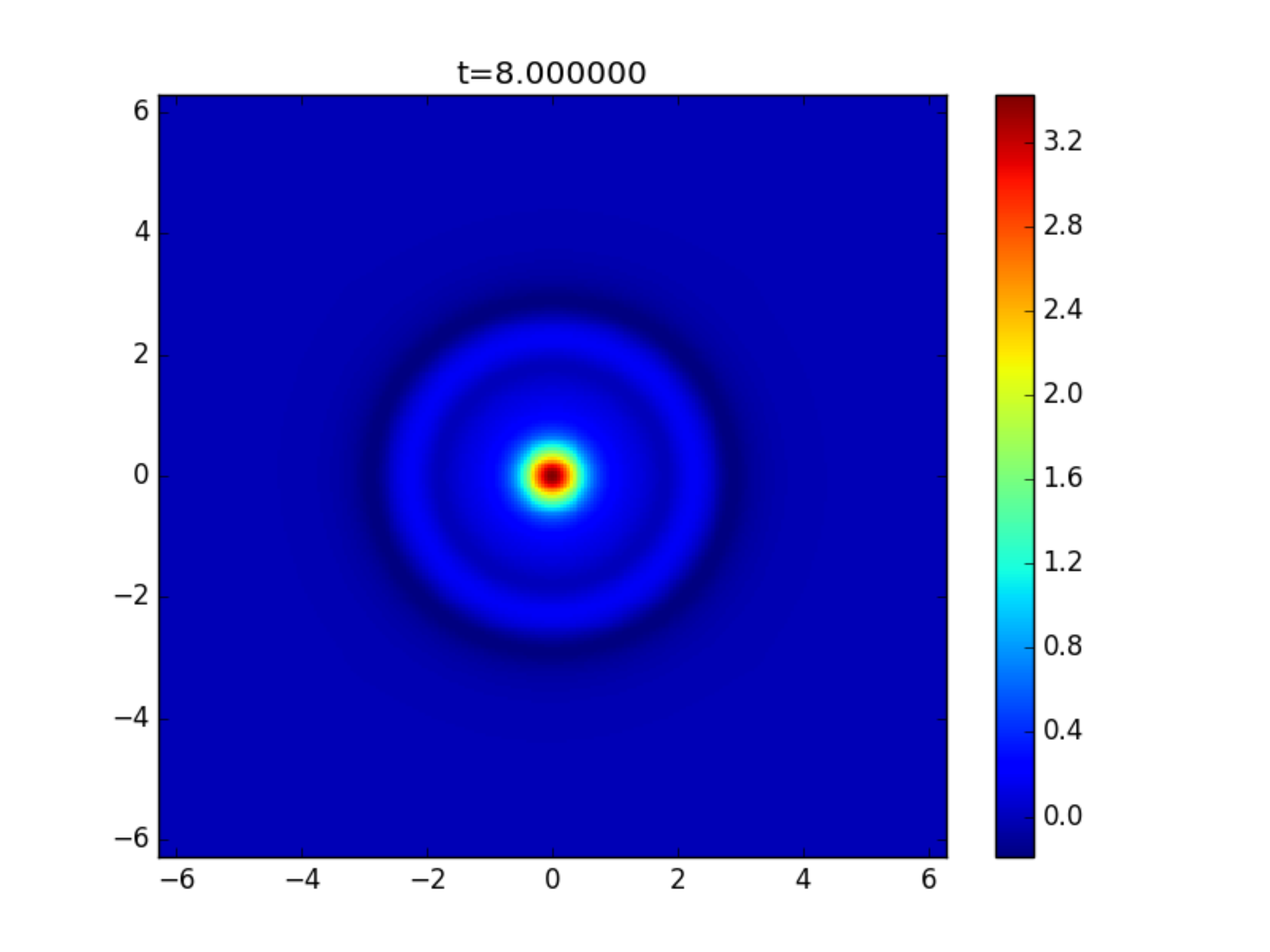} \\
       \includegraphics[width=5cm, height=4cm]{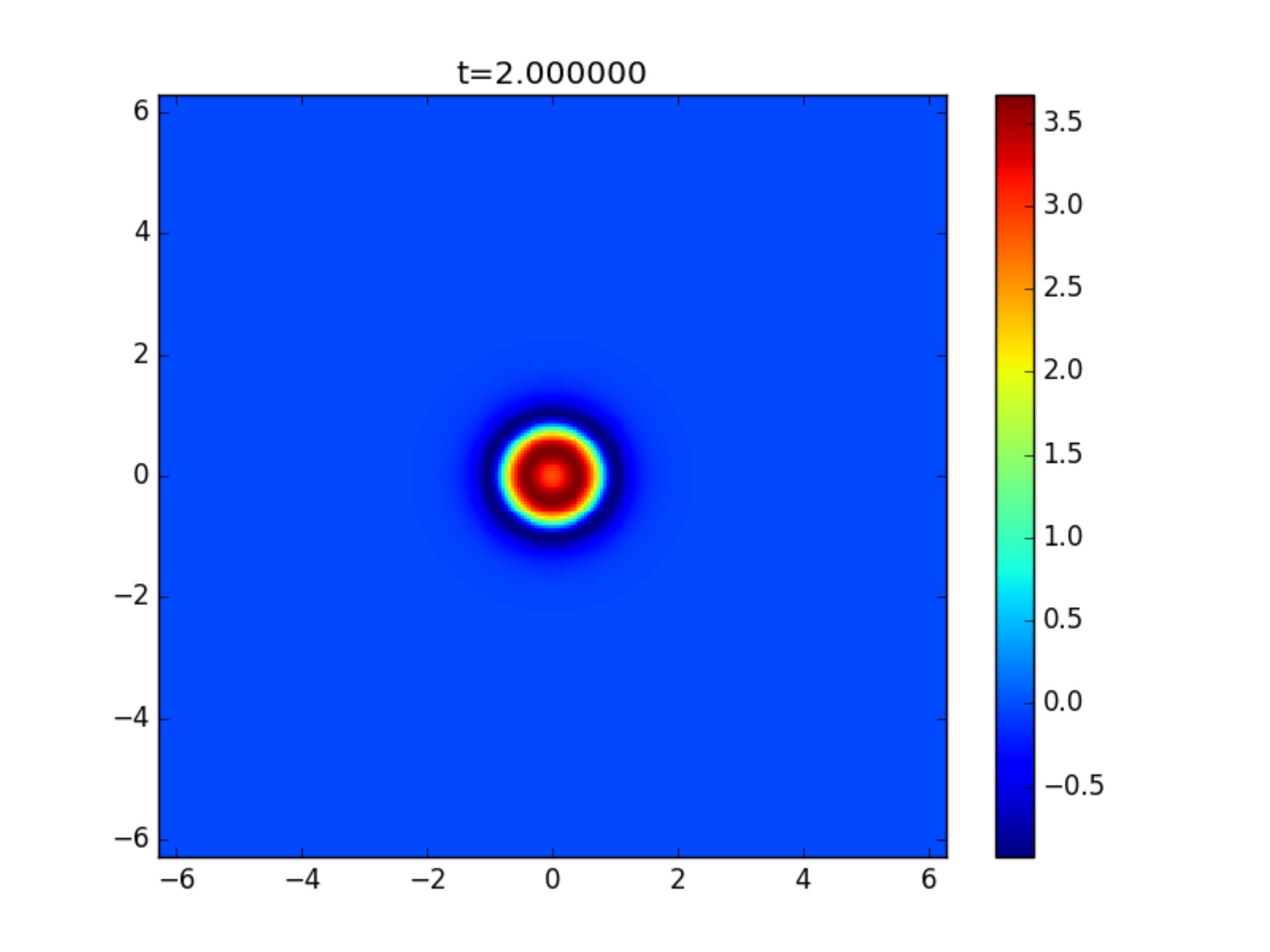} 
      \includegraphics[width=5cm, height=4cm]{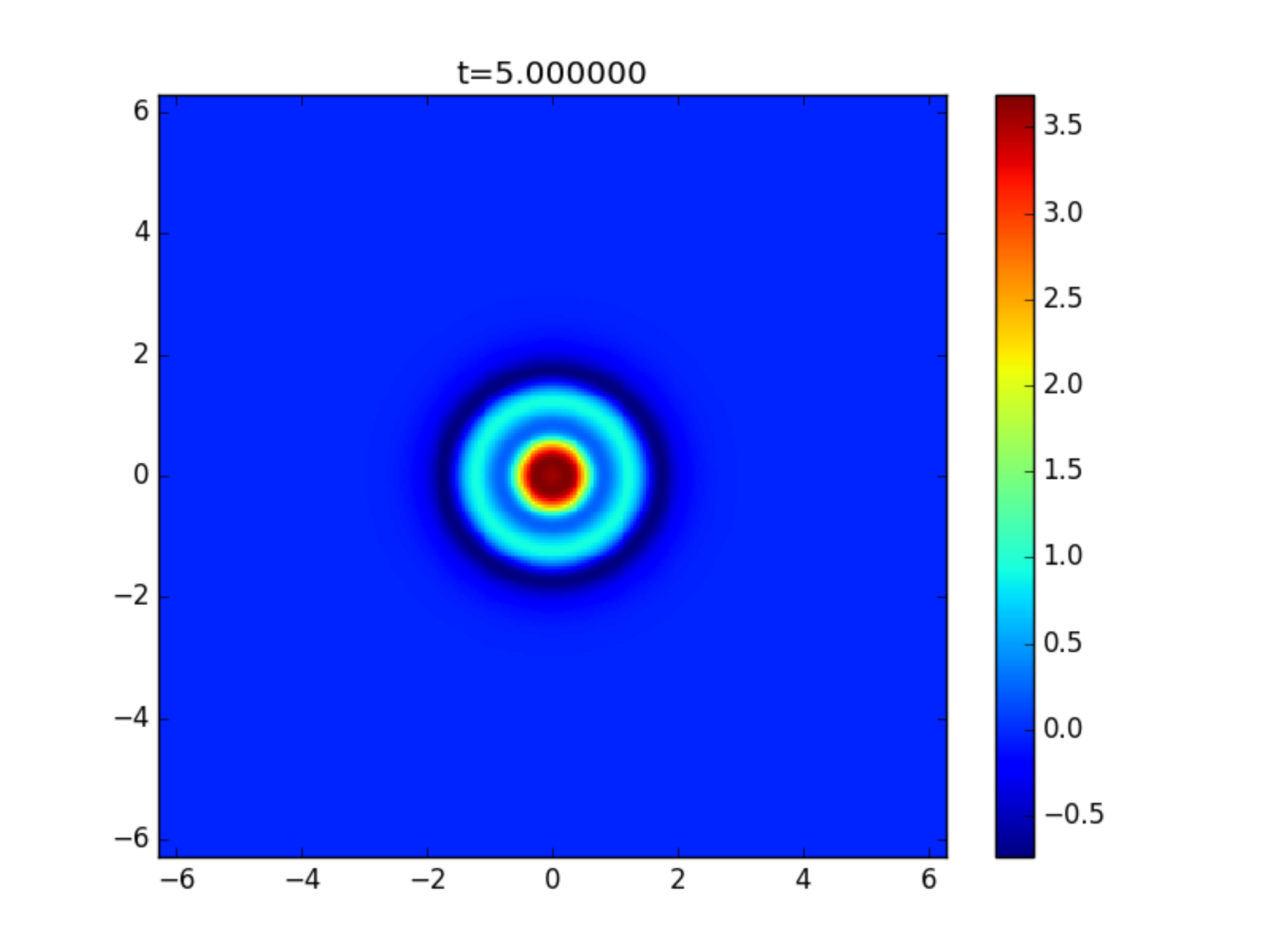}
       \includegraphics[width=5cm, height=4cm]{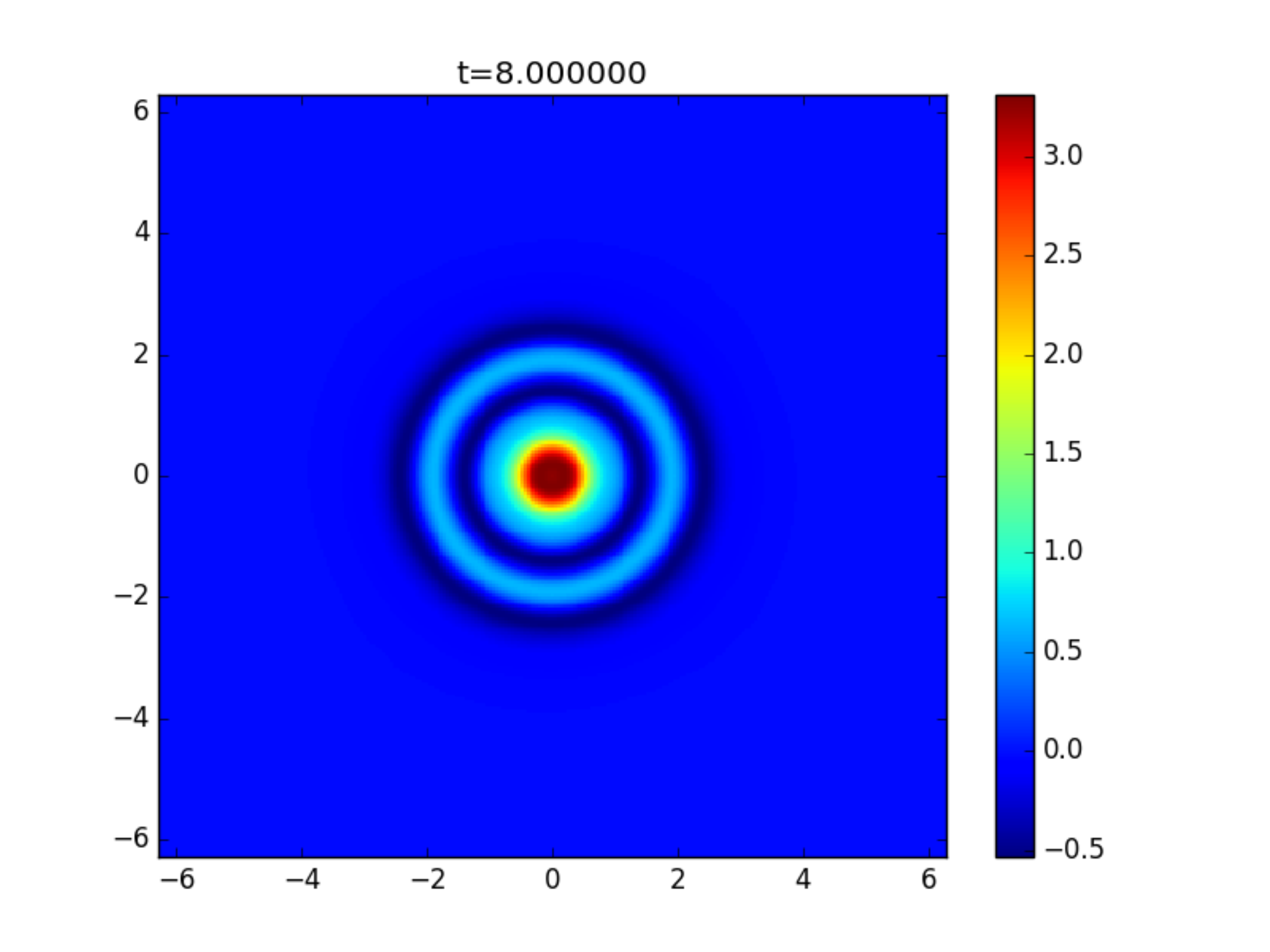}     
          \caption{Gaussian vorticity profile in $x-y$ plane for Reynolds number $R=2.0$, Mach number $M=0.4$, nd viscoelastic coefficient top($\tau_{m}=1$), middle ($\tau_{m}=5$), and bottom ($\tau_{m}=10$) }
             \label{fig-gaussian}
   \end{figure*}

\begin{figure}[h!]
       \centerline{    \includegraphics[scale=1]{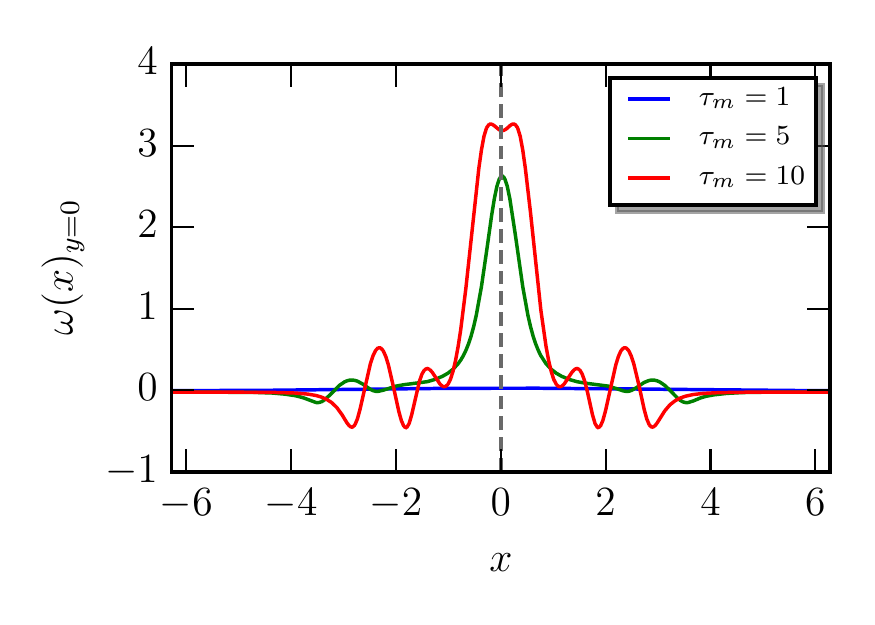} }
             \caption{Linear profile of spatial vorticity for $y=0$ axis with $R=2.0$, various $\tau_{m}$ and Mach number $M=0.4$ at time $t=10.0$ } 
  \label{fig-varytau}
\end{figure}

\begin{figure*}  
 \centerline{ 
  \includegraphics[width=5cm, height=4cm]{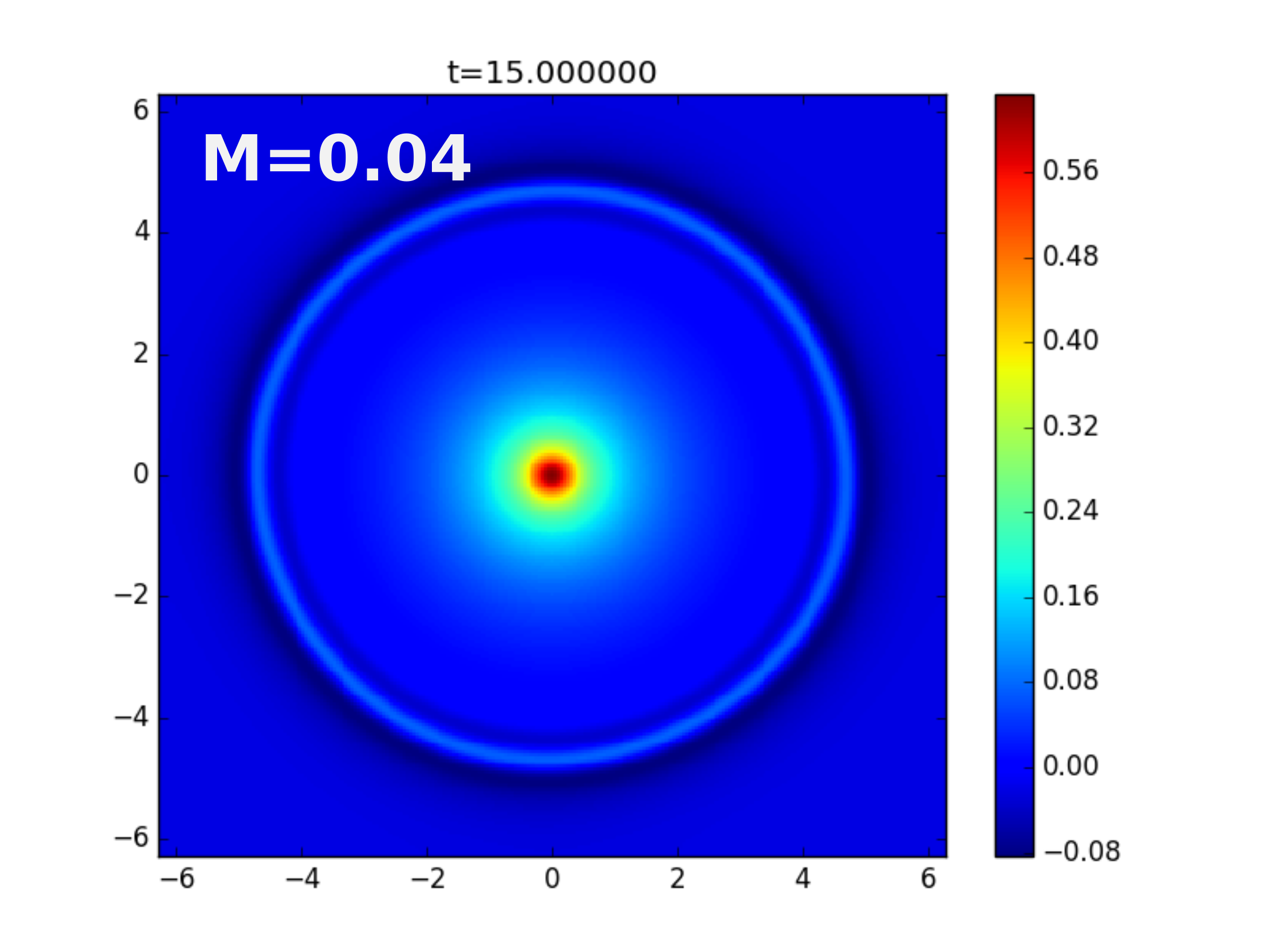} 
      \includegraphics[width=5cm, height=4cm]{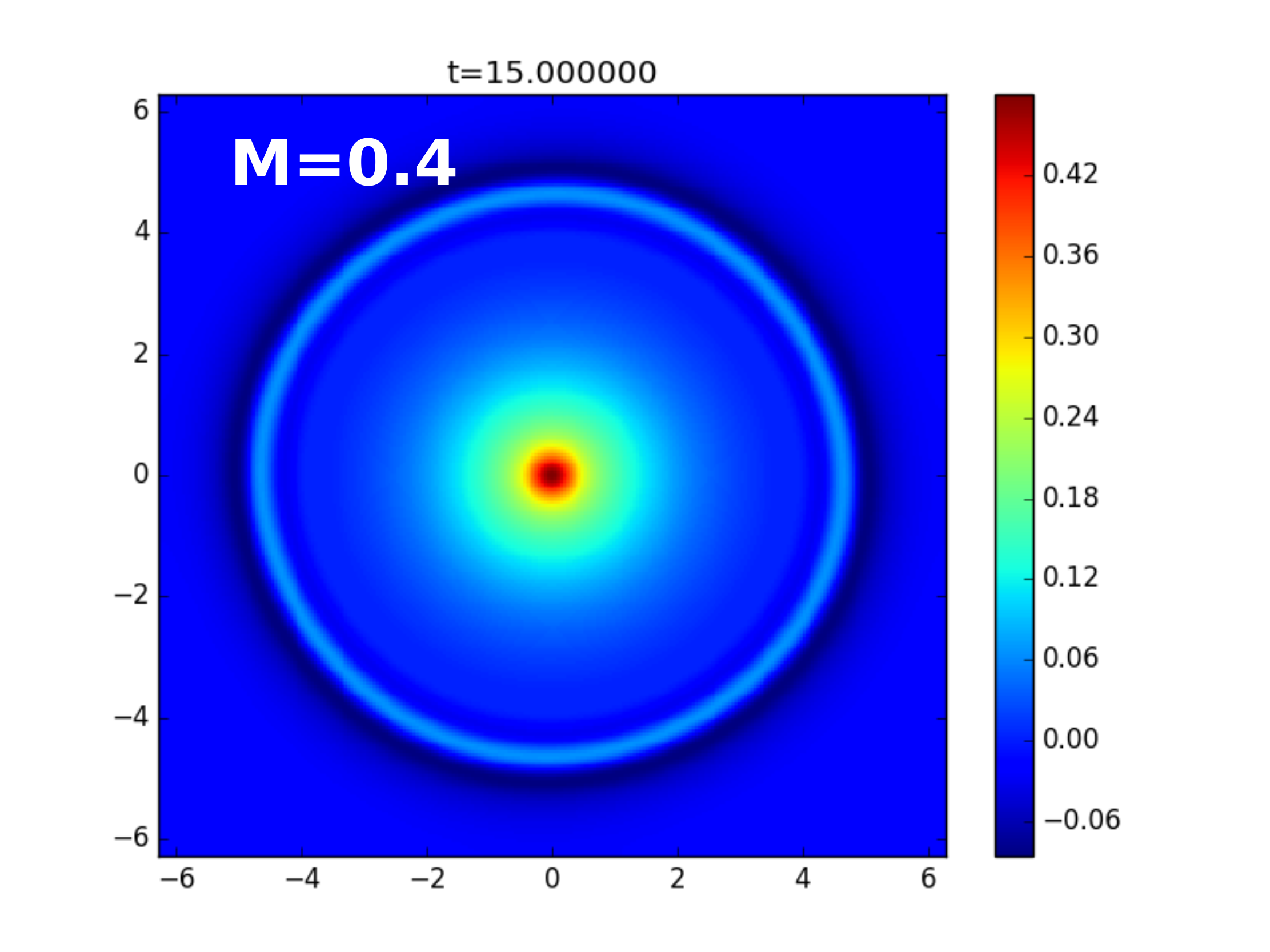}
       \includegraphics[width=5cm, height=4cm]{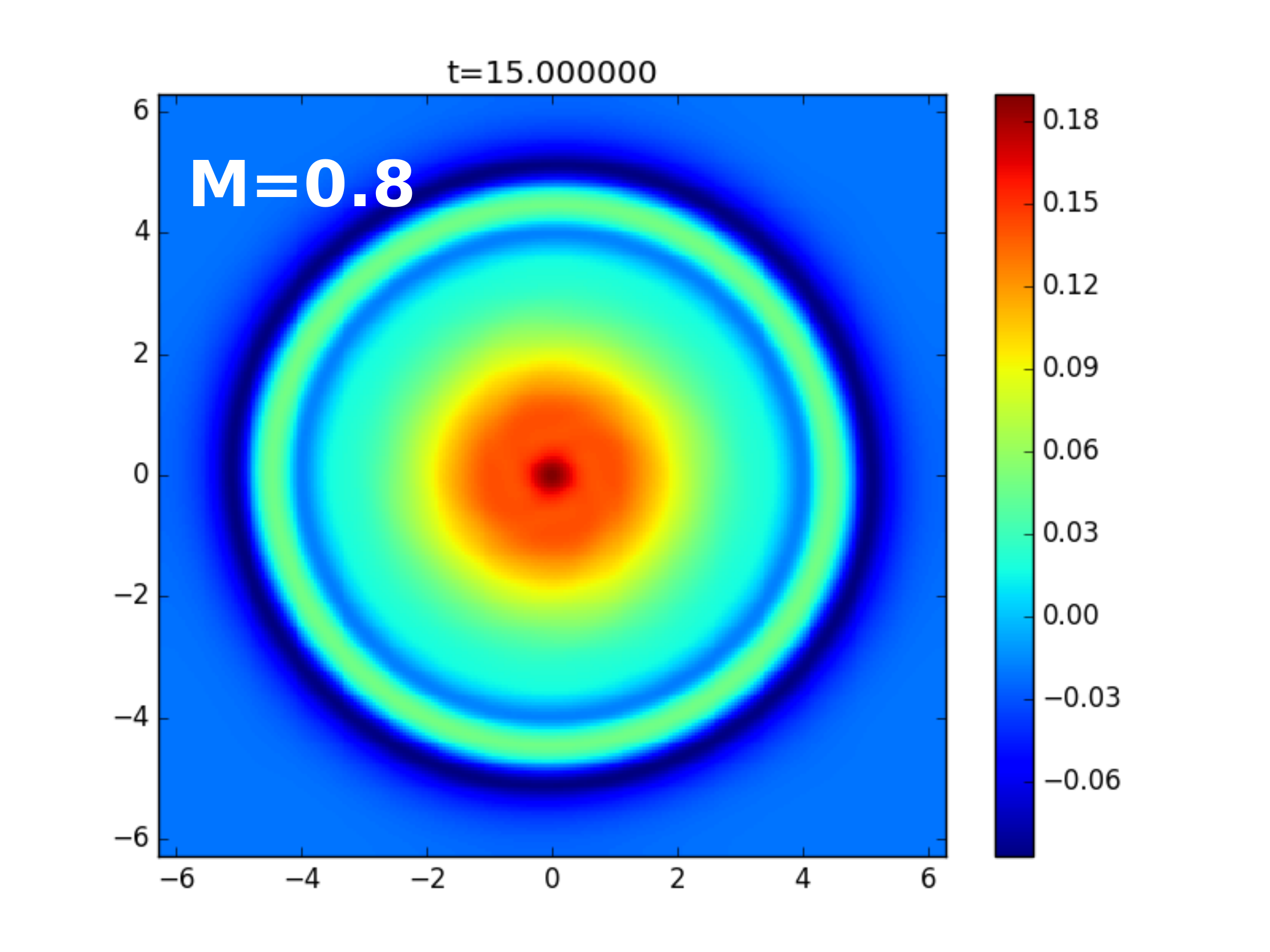} } 
            \caption{Gaussian vorticity profile in $x-y$ plane. Effect of compressibility for $M \neq 0, \tau_{m}=5.0, R=2.0$ at time $t=15.0$ }   
              \label{fig-machvary}   
           \end{figure*}

\begin{figure*}
\begin{center}
  
  \includegraphics[scale=0.6]{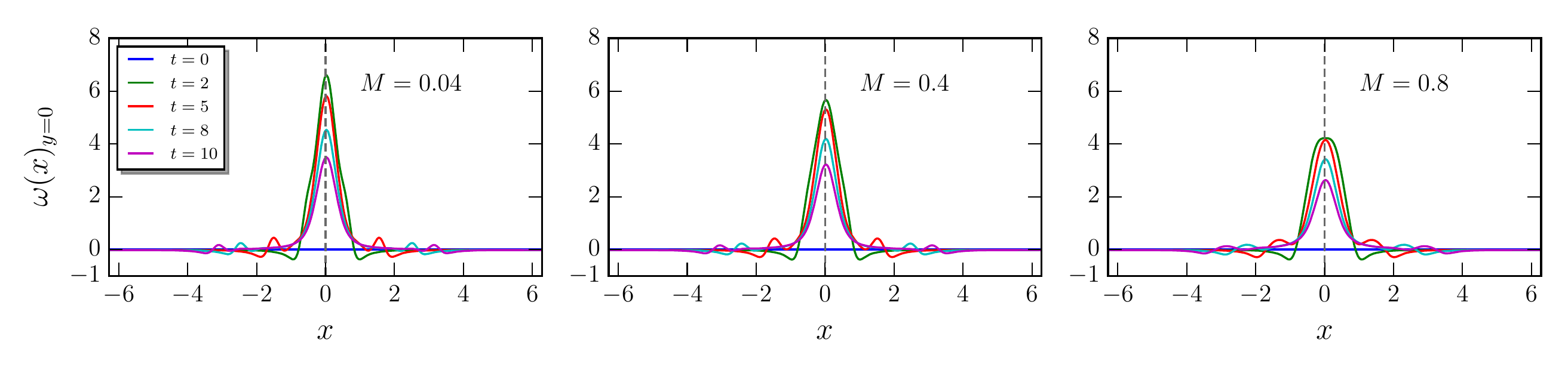}  
            \caption{Linear profile of spatial vorticity for $y=0$ axis with $R=2.0$, various Mach number $M$ and viscoelastic-coefficent $\tau_{m}=5.0$.  }   
              \label{fig-machvary_vorti}   
              \end{center}
           \end{figure*}

\begin{figure}[h!]
           \includegraphics[scale=1]{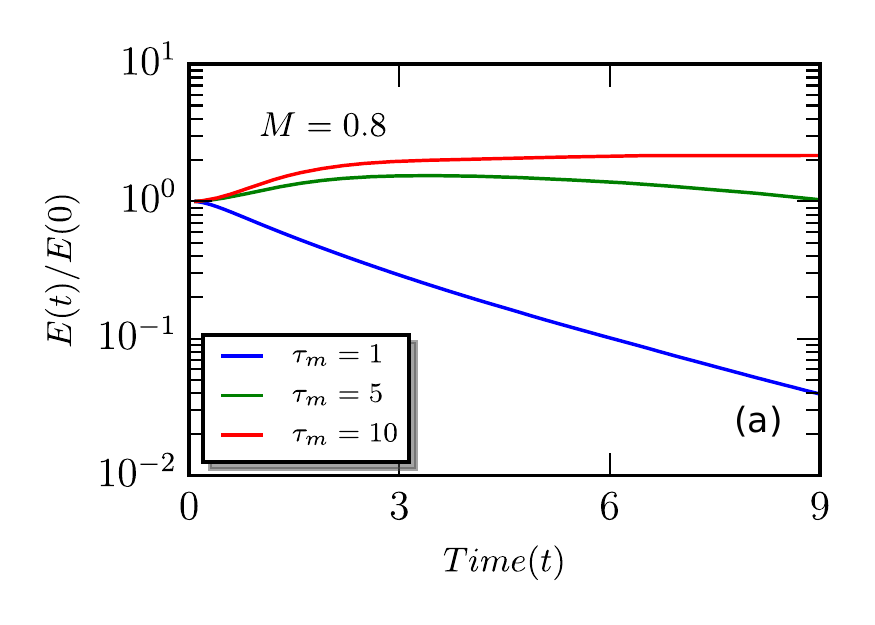} 
           \includegraphics[scale=1]{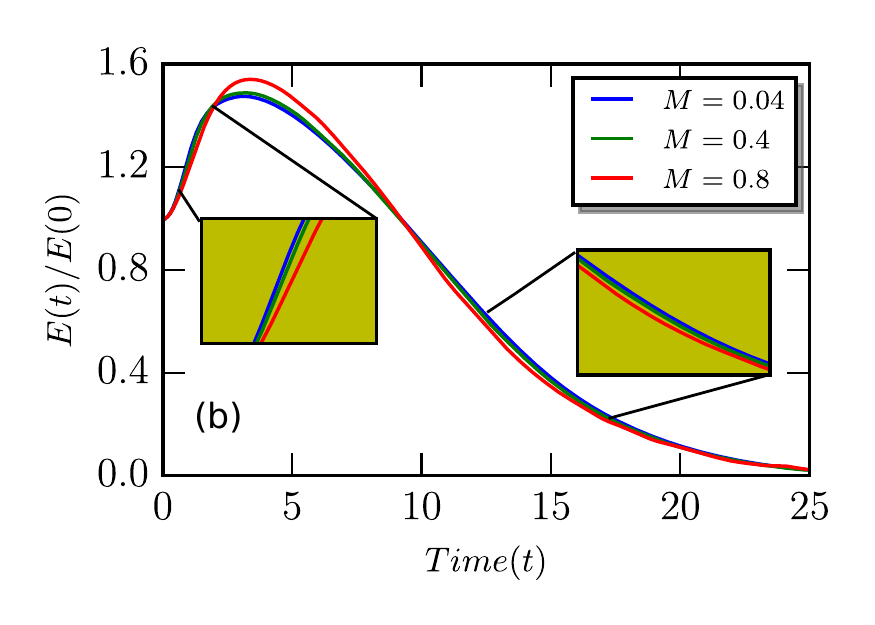} 
             \caption{(a) Total Kinetic energy for various $\tau_{m}$, $R=2$, and $M=0.8$, (b) Total Kinetic energy for various Mach number $M$ and $\tau_{m}=5.0$. } 
  \label{fig-ke}
\end{figure}

In the  present study the localized Gaussian vortex $\omega=\omega_{0}e^{-(x^{2}+y^{2})}$, where $\omega_{0}= 5.0$ has been considered as initial input profile.
Emergence and propagation of transverse shear
wave towards periodic boundaries have been observed. In Fig.~\ref{fig-gaussian}, the evolution of Gaussian vorticity profile has been plotted for various values of viscoelastic-coefficient ($\tau_{m}$). It is evident from the figure that the viscoelasticity delayed the generation of transverse shear. To study the effect of viscoelasticity over the vorticity, we plot in Fig.~\ref{fig-varytau} the linear profile of spatial vorticity for $y=0$ axes with $R=2.0$, various $\tau_{m}$ and Mach number $M=0.4$ at time $t=10.0$. Viscoelasticity suppresses the dispersion of vorticity and vortex are more concentrated at the center.  We also study the effect of compressibility in  Fig.\ref{fig-machvary} and Fig.\ref{fig-machvary_vorti}. We find compressibility does not show any such effect over the outer-core of the vortex, however, the magnitude of vorticity inside the outer-region has changed, which is more clear in Fig.\ref{fig-machvary_vorti}. The total kinetic energy of the system has also been studied for various values of $\tau_m$ and the fixed value of Mach number $M=0.8$. In Fig.\ref{fig-ke}(a), kinetic energy for various values of $\tau_m$ shows that viscoelasticity reduces the dissipation due to viscosity. For the lower $\tau_{m}$ value viscous dissipation dominates and vortex energy decays as time evolves, however, for higher viscoelasticiy $\tau_m=5$ and $\tau_m=8$ cases with nonlinear saturation has also been observed.   \\

Fig.\ref{fig-ke}(b) is for the total kinetic energy with time plot for various values of Mach number for  $\tau_{m}=5$. In this figure, it is evident that the compressibilty does not show any significant changes.   However, for a larger value, a  nonlinear saturation occurs for a short time $t=2-6$ where the viscous dissipation is compensated by nonlinear terms. After time $t=6$ the viscous dissipation dominates. Small effect of compressibilty has been observed (see the insert zoom plot in Fig.~\ref{fig-ke}(b)) before and after nonlinear saturation regime. \\

In the next section, we study the effect of viscoelasticity and compressibility on the single and multiple rotational vortices.


\subsection{Single rotating vortex: Rankine vortex}
\begin{figure*}
\includegraphics[width=5cm, height=4cm]{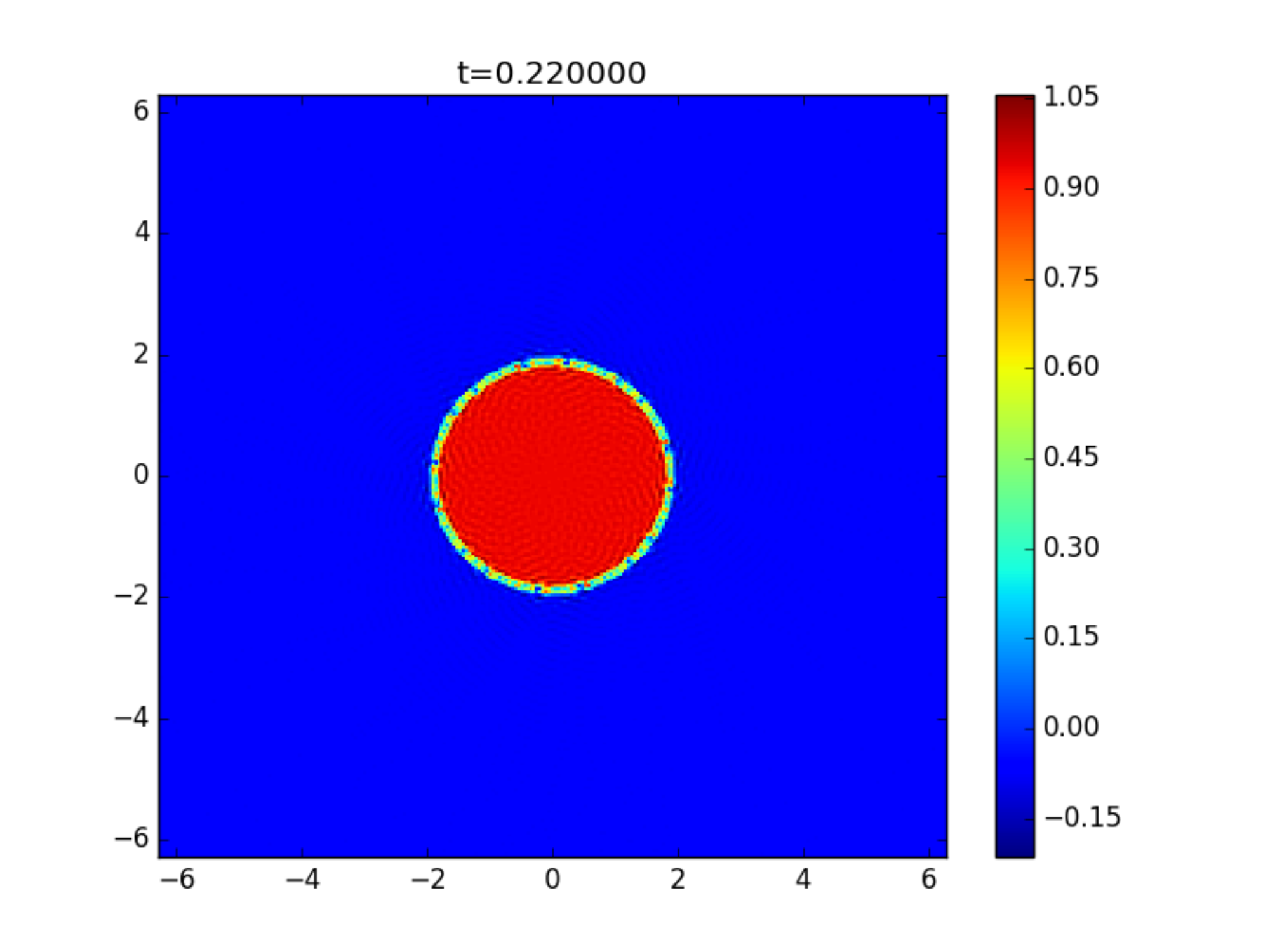}
\includegraphics[width=5cm, height=4cm]{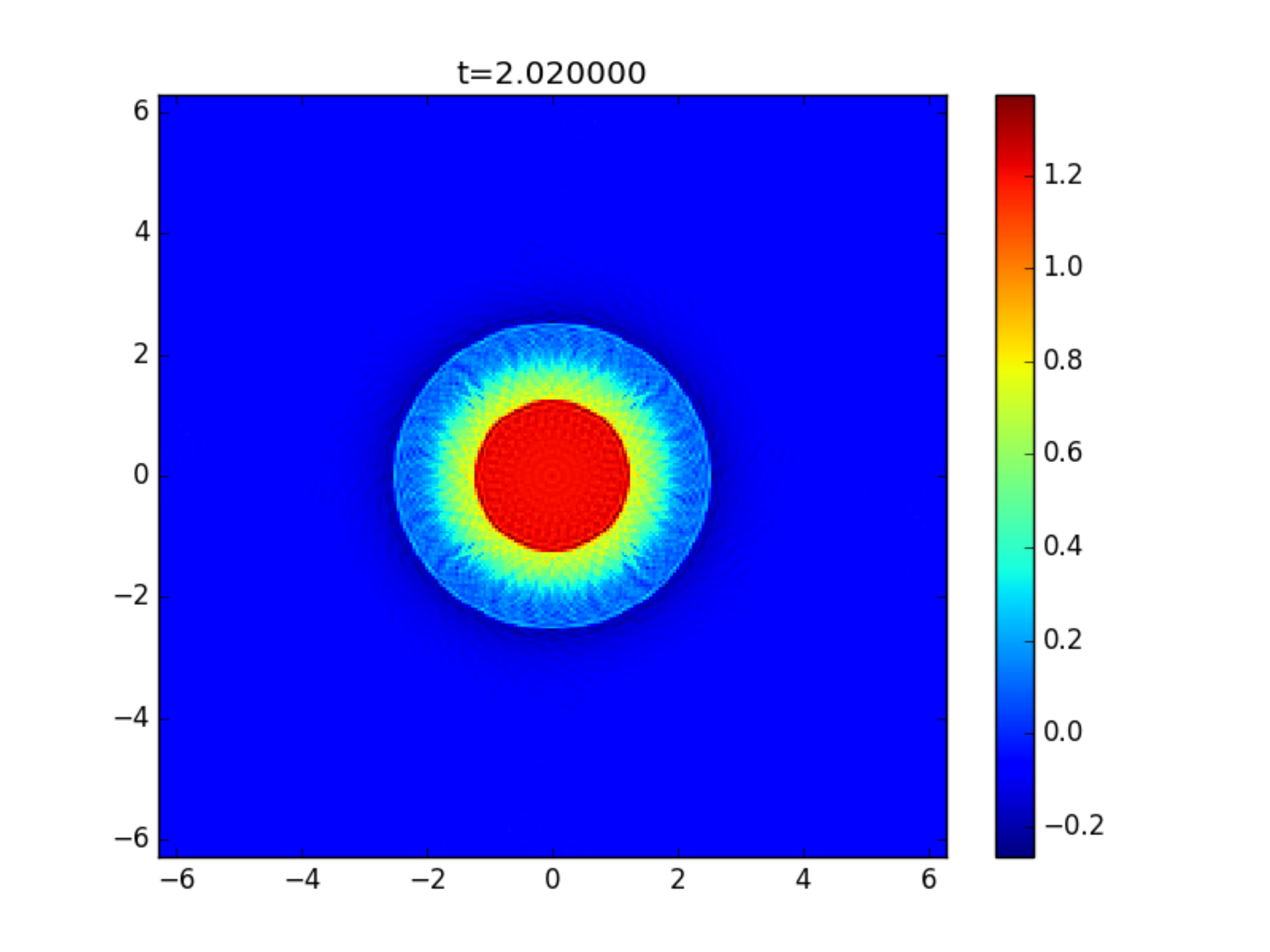}
\includegraphics[width=5cm, height=4cm]{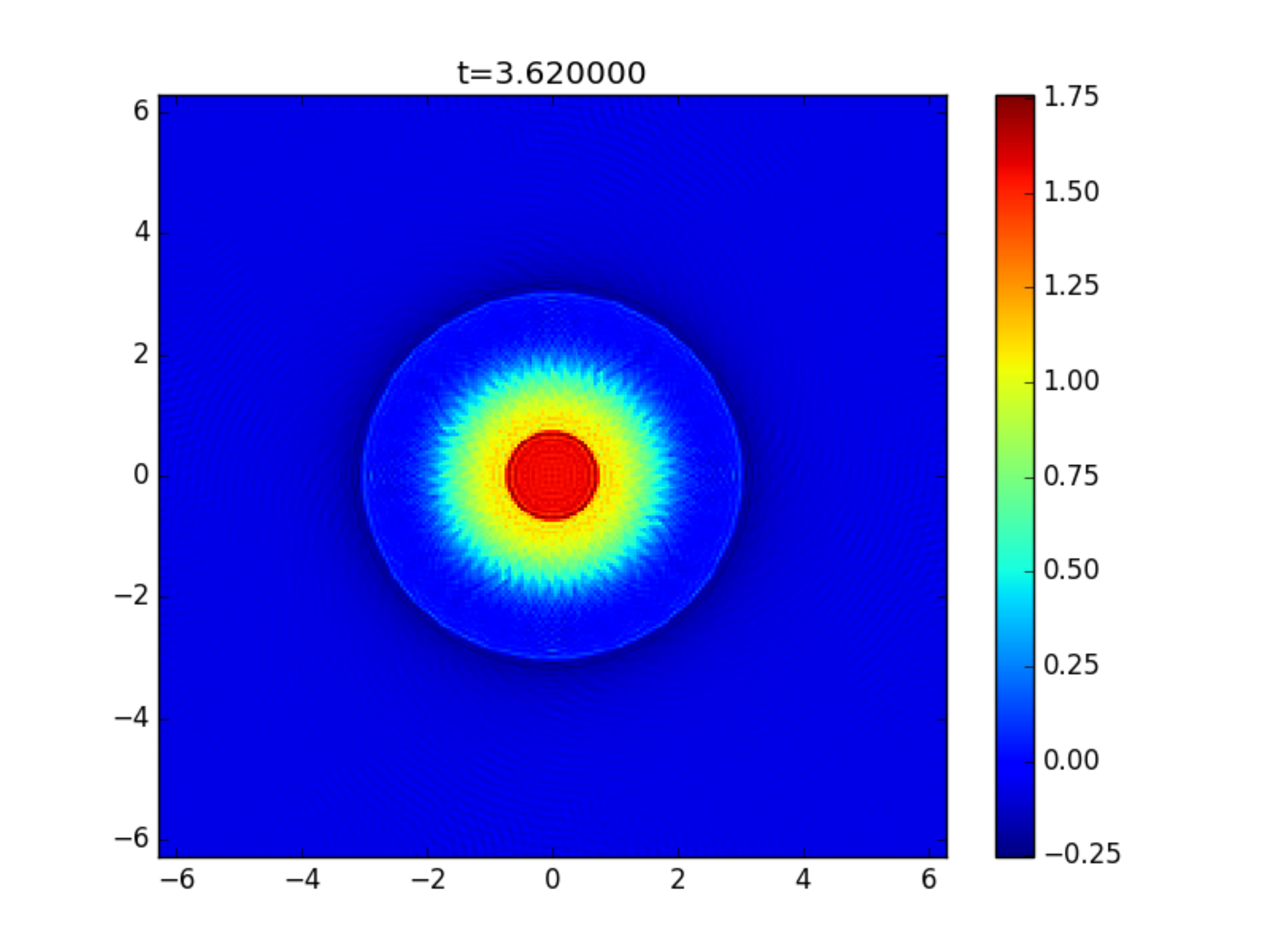}\\
\includegraphics[width=5cm, height=4cm]{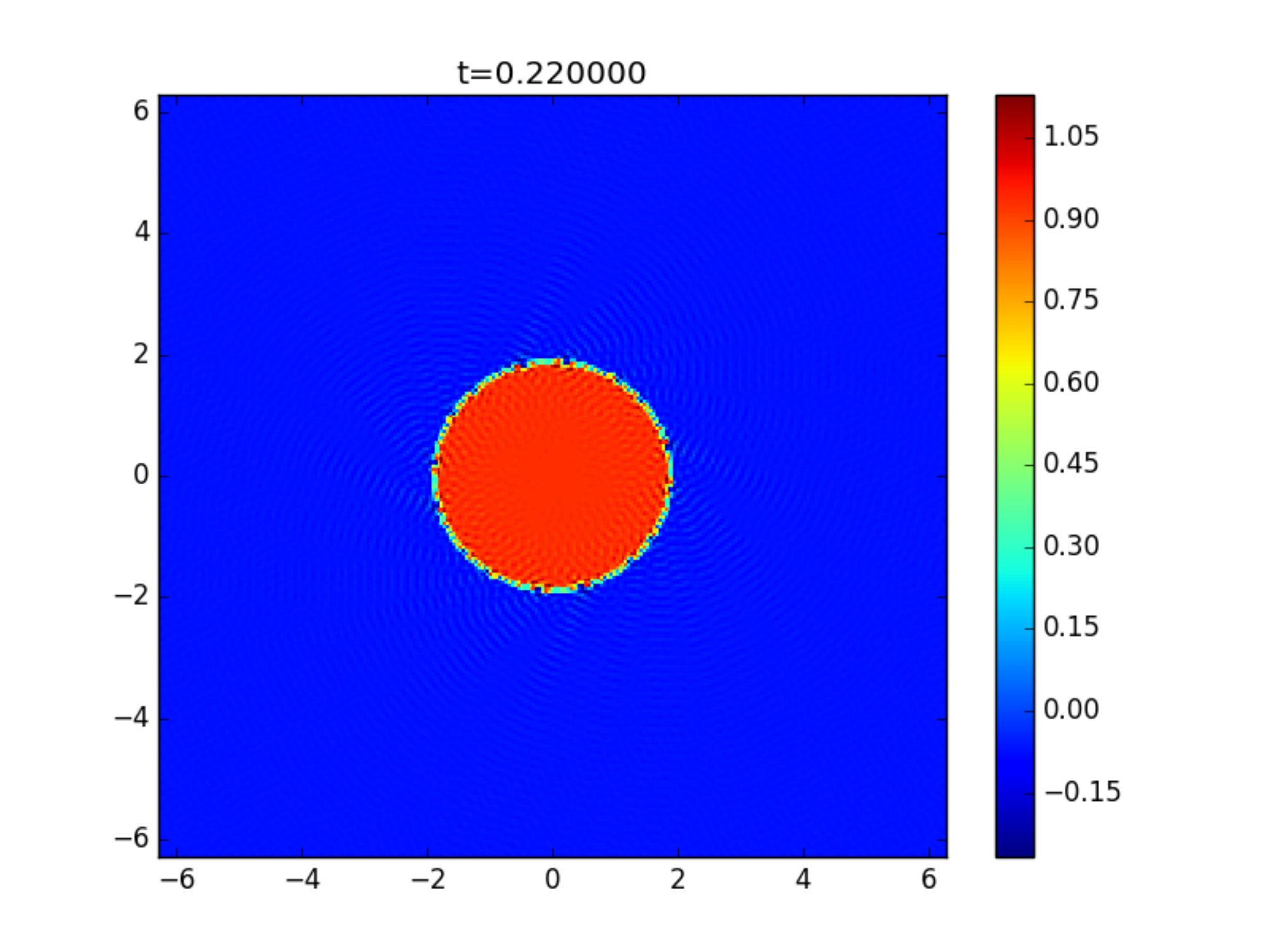} 
\includegraphics[width=5cm, height=4cm]{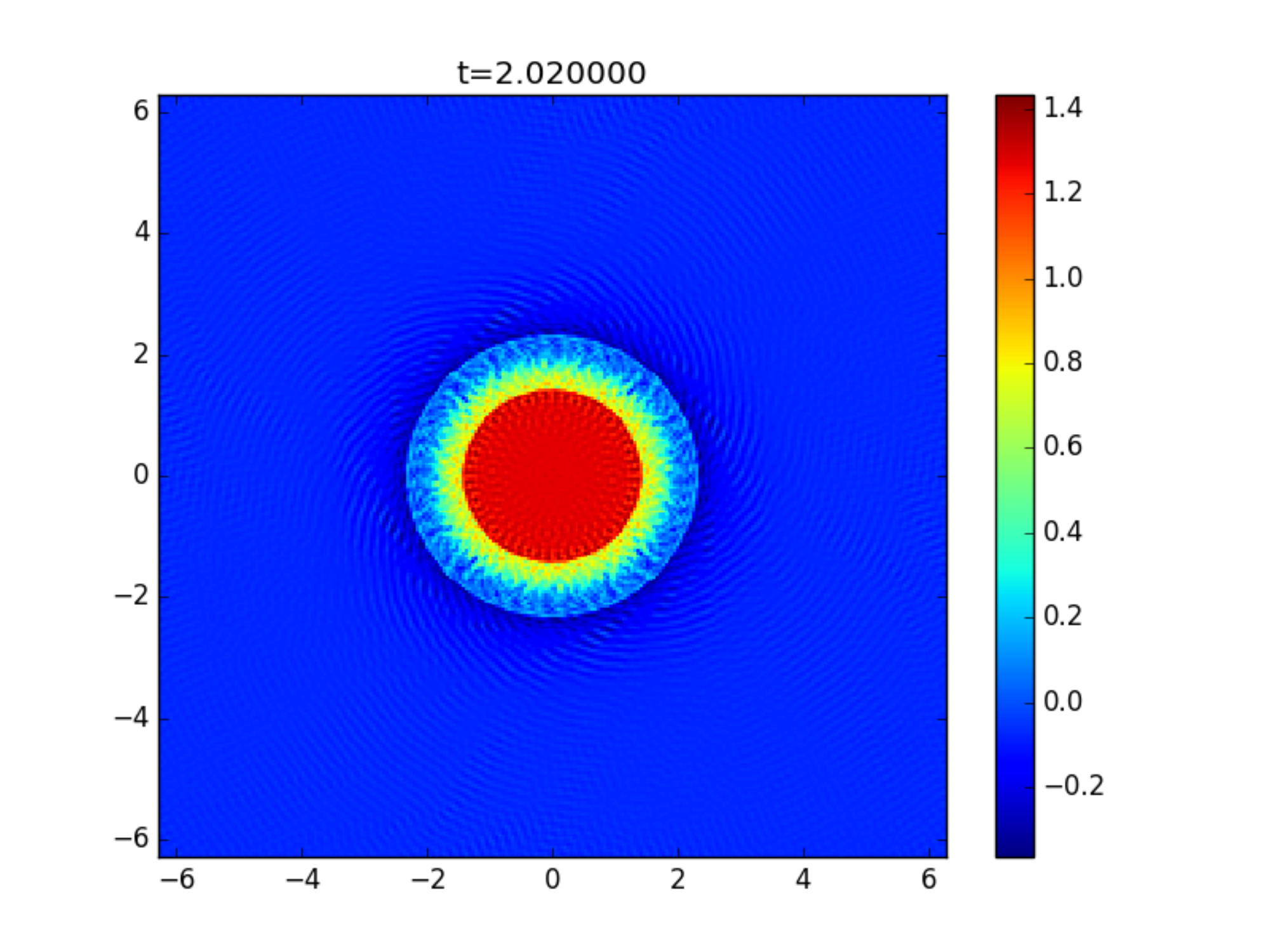}
\includegraphics[width=5cm, height=4cm]{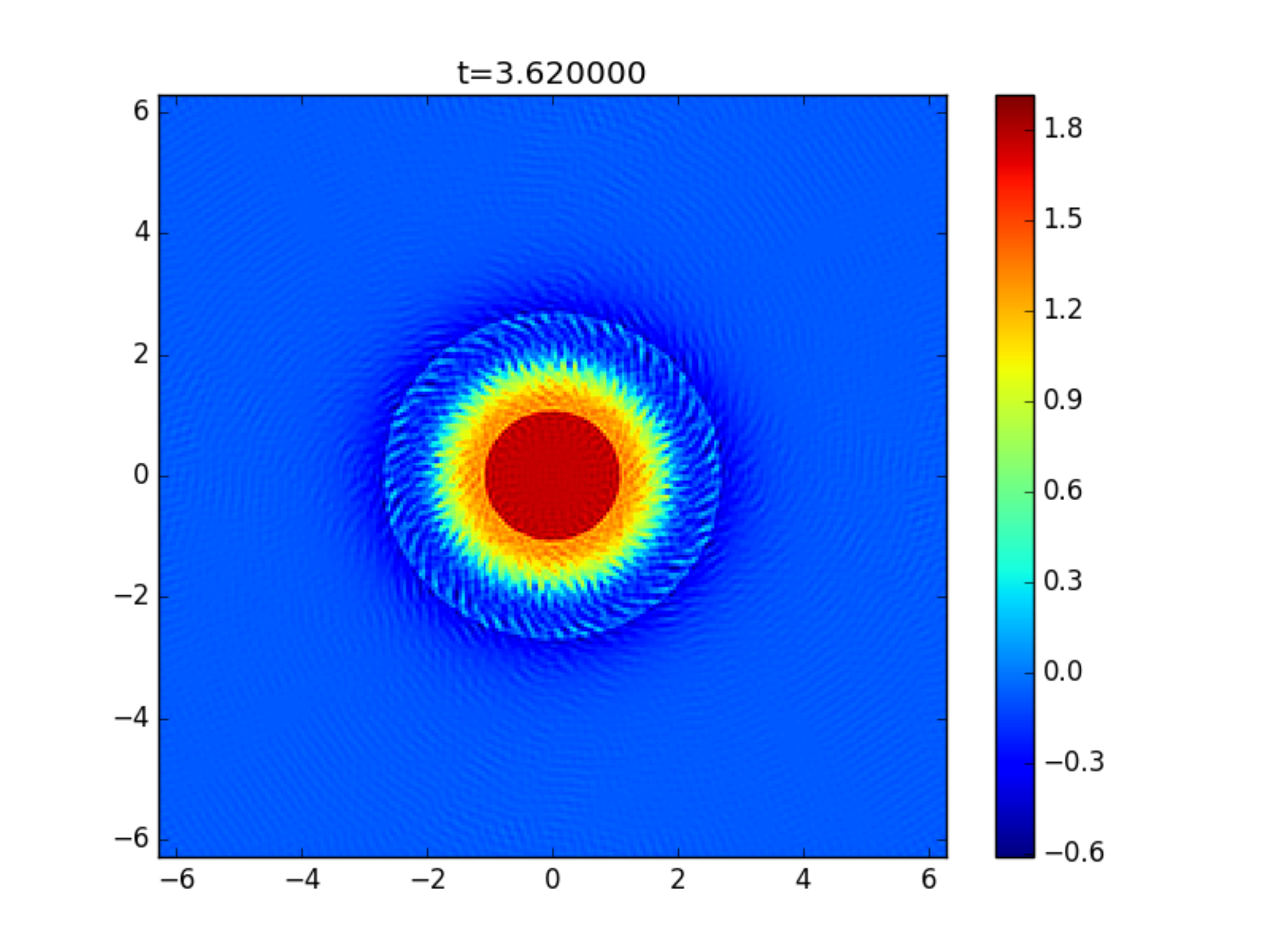}  
\caption{(Color online) Time evolution of a Rankine vortex for $\tau_m =$ $5$ (upper pannel) and $10$ (lower pannel) with $C_s^2 = 354.945$ and $\omega_0 = 1$.}
\label{snap_rankine}
\end{figure*}

A single patch vortex (also known as ``Rankine vortex") is a solution of a fluid without memory effects. Thus when time evolved, the vortex remains stationary. For a dusty plasma medium, a Rankine vortex gets de-stabilized by itself and generates several patterns as the time is evolved \cite{vikram:2014} due to the effect of viscoelasticity. In the following,  we provide some snapshots of the evolution of a single Rankine vortex in a viscoelastic fluid Fig.~\ref{snap_rankine}.

The effect of viscoelasticity is also found in the case of a translational motion of a point vortex. Any point vortex moves in time keeping an impression of the previous position of the vortex itself. Thus the merger process of a Rankine vortex and some point-like vortices in a strongly coupled dusty plasma become completely different compared to a neutral fluid merger process. Therefore in the next section, we study a pre-arranged set of vortices of two different radii and observe the merger process.

\subsection{Multiple rotating vortices}
\begin{figure}[h!]
\centerline{\includegraphics[scale=0.43]{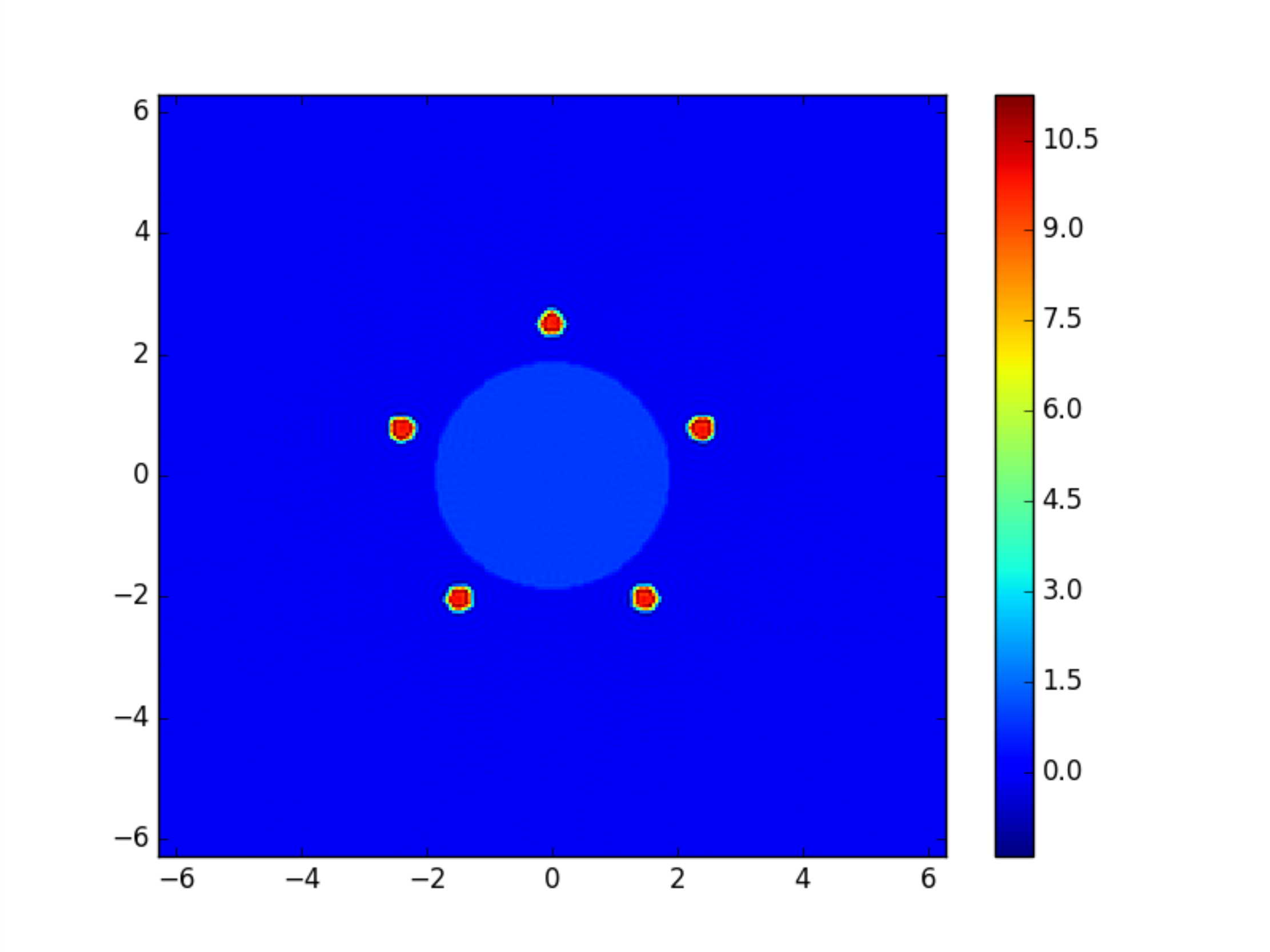}}
\caption{Vorticity contour plot of the initial profile. The pseudocolor bar indicates the strength of vorticity.}
\label{frame0}
\end{figure}

We start with a central patch vortex of radius $R_{pa}$ and vorticity $\omega_{pa}$ and $N_{pv}$ number of pre-arranged equispaced point-like vortices of vorticity $\omega_{pv}$ around the patch vortex at a distance $R_{pv}$ from the center of the patch vortex. The vorticity contour plot of this configuration is shown in Fig.~\ref{frame0}. From previous study \cite{ganesh:2002, deem:1978, lansky:1997, durkin:2000} it is known that $N_{pv}$ should be chosen below a critical value ($N_{pv} = 7$) to have persistent vortex hole pattern. We also note $\frac{R_{pv}}{R_{pa}} = \sqrt{\frac{N_{pv}}{N_{pv}-1}}$ from earlier studies \cite{ganesh:2002, deem:1978, lansky:1997} and choose our parameters such that the frequencies generated do not get affected by  choice of the ratio $\frac{R_{pv}}{R_{pa}}$ and still the point-like vortices can excite the Kelvin waves on the surface of the patch vortex. \\

Using the above initial profiles, we proceed to simulate the vortex merger problem discussed above using AG-Spect code. We choose the system size to be big enough so that, the finite size effect can be neglected. We keep the value of $\nu$ small enough such that, we can isolate the effect of viscoelasticity. For numerical purpose, we further choose a finite radius of each point vortices as $d_{pv} = 0.032 \frac{L}{2}$. We define the patch turnover time as $\tau_D = \frac{2 \pi}{\omega_{pa}} = \frac{2 \pi}{1} = 2 \pi$. For further check, we compare our results with identical parameter set (provided in Table \ref{parameter}) using another well-benchmarked code C-FD-2D \cite{rupak:2018, rupak:2018:IEEE} for a neutral fluid scenario having $\tau_m = 0$. The code C-FD-2D is an OpenMP parallel, pseudo-spectral code capable of simulating weakly compressible two dimensional Navier-Stokes equations in conservative form, using Cartesian co-ordinates with periodic boundary conditions. The time solvers of C-FD-2D are explicit and the convergence of the nonlinear test results are checked with Runge-Kutta 4, Predictor-Corrector and Adams-Bashforth algorithms. The de-aliasing technique used for C-FD-2D to remove spurious Fourier components during the evaluation of nonlinear terms is zero-padding method. We evolve the prearranged vortices in a system of area $(4\pi)^2$ with a resolution of $256^2$ grids with time-steps of $10^{-5}$. We obtain the identical result with another run with grid size $512^2$. This indicates the numerical convergence of our results. For the rest of the paper, unless otherwise indicated, we keep the grid resolution at $256^2$. The parameter regime for which we compare the results of the two codes (AG-Spect and C-FD-2D) are given in the following table.\\
\begin{table}[h!]
\centering
\begin{tabular}{ |c|c|c|c|c|c|c|c|c| }
 \hline
 $N_x = N_y$ & $L_x = L_y$ & $dt$ & $R_{pa}$ & $R_{pv}$ & $N_{pv}$ & $\omega_{pa}$ & $\omega_{pv}$ & $n_{d_0}$\\
 \hline
 ~ & ~ & ~ & ~ & ~ & ~ & ~ & ~ &\\
 $256$ & $4 \pi$ & $10^{-5}$ & $0.3\frac{L}{2}$ & $0.4 \frac{L}{2}$ & $5$ & 1 & $10$ & $1$ \\
 ~ & ~ & ~ & ~ & ~ & ~ & ~ & ~ &\\ 
 \hline
\end{tabular}
\caption{Parameter details for the benchmark of the code C-FD-2D with AG-Spect. These parameters are kept identical throughout this report unless stated otherwise.}
\label{parameter}
\end{table}

\begin{figure}[h!]
\begin{center}
\centerline{\includegraphics[scale=1]{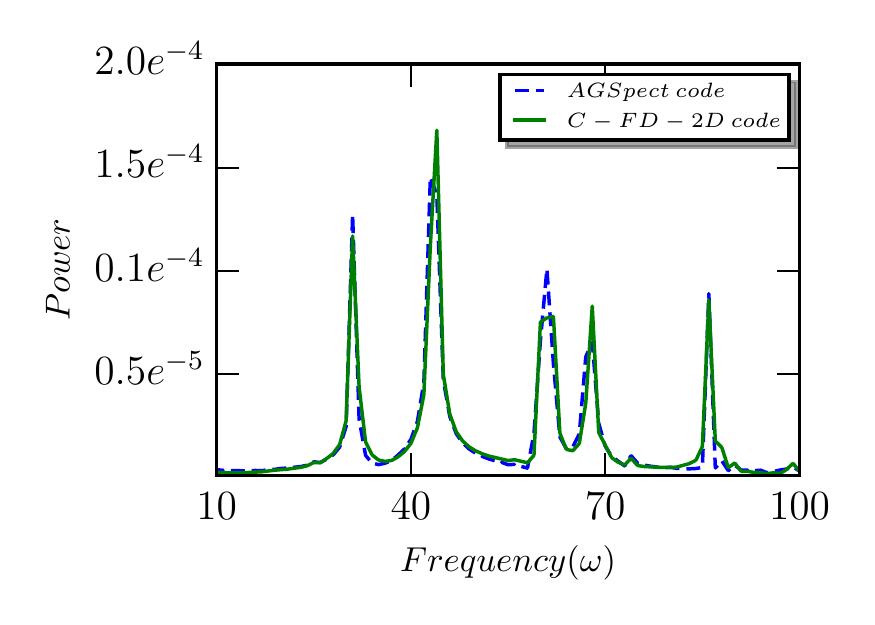}}
\caption{(Color online) Comparison of natural frequency and its harmonics as well as its beats with its harmonics from two different codes, AGSpect (red solid line) with $\tau_m = 10^{-4}$ and C-FD-2D (blue dotted line) with $\tau_m = 0$ and $M = 0.4$ for a neutral fluid problem.}
\label{comparison}
\end{center}
\end{figure}

We found the results obtained from the two independent codes both in the incompressible ($M = 0.04$) and compressible ($M = 0.4$) limits to be quite satisfactory [Fig~\ref{comparison}].\\

We keep the above set of parameters identical throughout our paper unless stated otherwise. Next we vary only viscoelasticity via $\tau_m$ from $10^{-4}$ to $5$, keeping the transverse speed constant ($\sqrt{\frac{\nu}{\tau_m}} = 1.412$) and get the following observations.


The effect of visco-elasticity is also found in the case of  translational motion of a point vortex. Any point vortex moves in time keeping an impression of the previous position of the vortex itself. Thus the merger process of a Rankine vortex and some point-like vortices in a strongly coupled dusty plasma becomes completely different from a neutral fluid merger process. Fig.~\ref{snap_tau_m_incomp} shows some examples of this effect.

\subsubsection{Results for small memory}
\begin{figure*}[h]
\begin{center}
\includegraphics[scale=1]{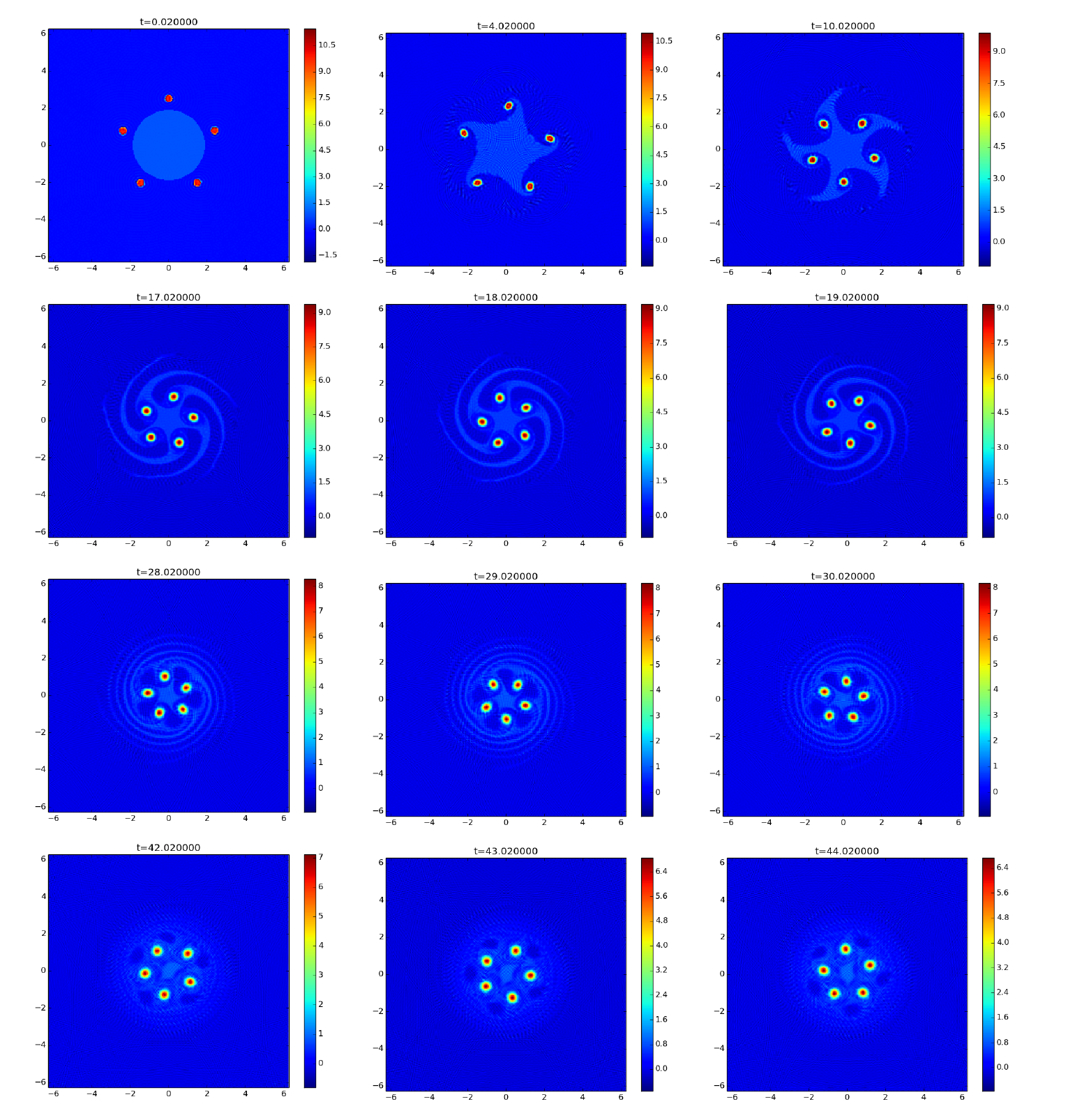}

\caption{(Color online) Time evolution of the prearranged vortex merger with $N_{pv} = 5$, $M = 0.04$, $\tau_m = 10^{-3}$ and $\sqrt{\frac{\nu}{\tau_m}} = 0.447$.}
\label{snap_no_memory}
\end{center}
\end{figure*}

$\tau_m \rightarrow 0$ limit corresponds to  fluid without the effect of viscoelasticity. We time evolve the above set of equations with the pre-arranged structure of vortices and observe the merging quite similar to what was observed earlier for a neutral fluid or a pure electron plasma\cite{ganesh:2002}. Some snapshots of time evolution of such a pre-arranged vortex merger are given in Fig.~\ref{snap_no_memory} for $N_{pv} = 5$, $M = 0.04$ $\tau_m = 10^{-3}$ and $\sqrt{\frac{\nu}{\tau_m}} = 0.447$.

 Due to the very small but non-zero value of $\tau_m$ we observe weak transverse waves generated from the point-like vortices populating the medium. The waves propagate outwards from each of the point-like vortices and collide with each other at the region of patch vortex. The point-like vortices excite Kelvin waves at the boundary of the patch vortex. The Kelvin waves grow nonlinearly generating a ``V" state. Then the perturbation grows and generates filamentation in the form of extended ``fingers" from the patch vortex-the filaments ``wave-break" at later times. Kelvin waves, being {\it negative energy} waves and the interaction between the point-like vortices and the {\it nearly} circular patch vortex being {\it positive}, the point like vortices get dragged into the patch vortex. Further, they form a ``vortex crystal" along with some ``holes" (region of zero vorticity) within the distorted patch vortex. For $\tau_m = 0$ and incompressible neutral fluid, it has been shown that these vortex crystals are quite long-lived. \cite{ganesh:2002}\\

The time evolution of kinetic energy is found to be similar to what was observed earlier. The compressibility effect was studied in detail at $\tau_m = 10^{-4}$ and are found to match with the previous results \cite{rupak:2018}. In general, compressibility adds one more timescale - the sonic timescale - in the system. This delays the merger process and allows the vortex crystal to melt down. The point - patch vortex interaction gives rise to sonic waves that generate several frequencies in the system. These frequencies are found to be beats and the beats with the harmonics with the natural frequency of the system. A detailed analysis of these frequencies has been reported in an earlier communication. \cite{rupak:2018}


\subsubsection{Results for large memory}
The dynamics of the pre-arranged vortex pattern in a strongly coupled dusty plasma medium becomes completely different and gives birth of new patterns and process of evolution as time progresses when the medium possesses large viscoelasticity. Viscoelasticity introduces another new timescale in the system. This affects the whole dynamics of the system for a reasonably high value of $\tau_m \sim 10^{-2}$.

\begin{figure*}[h]
\begin{center}
\includegraphics[scale=1]{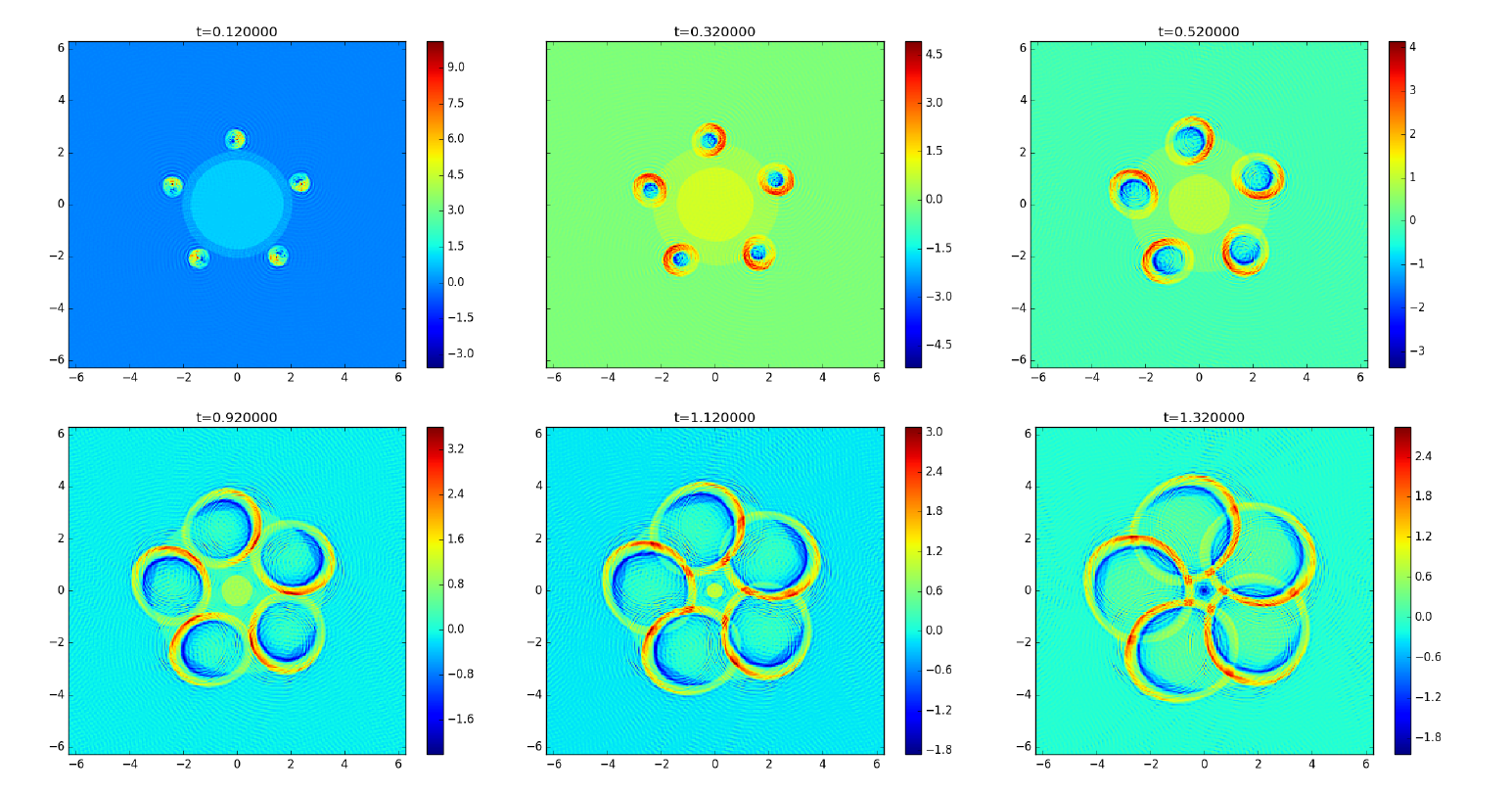}

\caption{(Color online) Time evolution of the prearranged vortex merger with $N_{pv} = 5$, $M = 0.04$, $\tau_m = 5$ and $\sqrt{\frac{\nu}{\tau_m}} = 1.412$.}
\label{snap_tau_m_incomp}
\end{center}
\end{figure*}

In case of a large value of $\tau_m$, the transverse waves arising out of the point-like vortices become more prominent and form  consecutive ring-like structures around the point vortex. The structures propagate in time becoming larger in radius and followed by other rings of smaller radii. The center of the ring moves with the time in the direction of flow, though it keeps an impression of the previous position of the point-like vortex. Thus the rings become elliptical in shape as time progresses. The ring from one point-like vortex collides with another formed from another point-like vortex. Some patterns of such merger process are provided in Fig.~\ref{snap_tau_m_incomp}. 

\begin{figure*}[h!]
\includegraphics[scale=1.0]{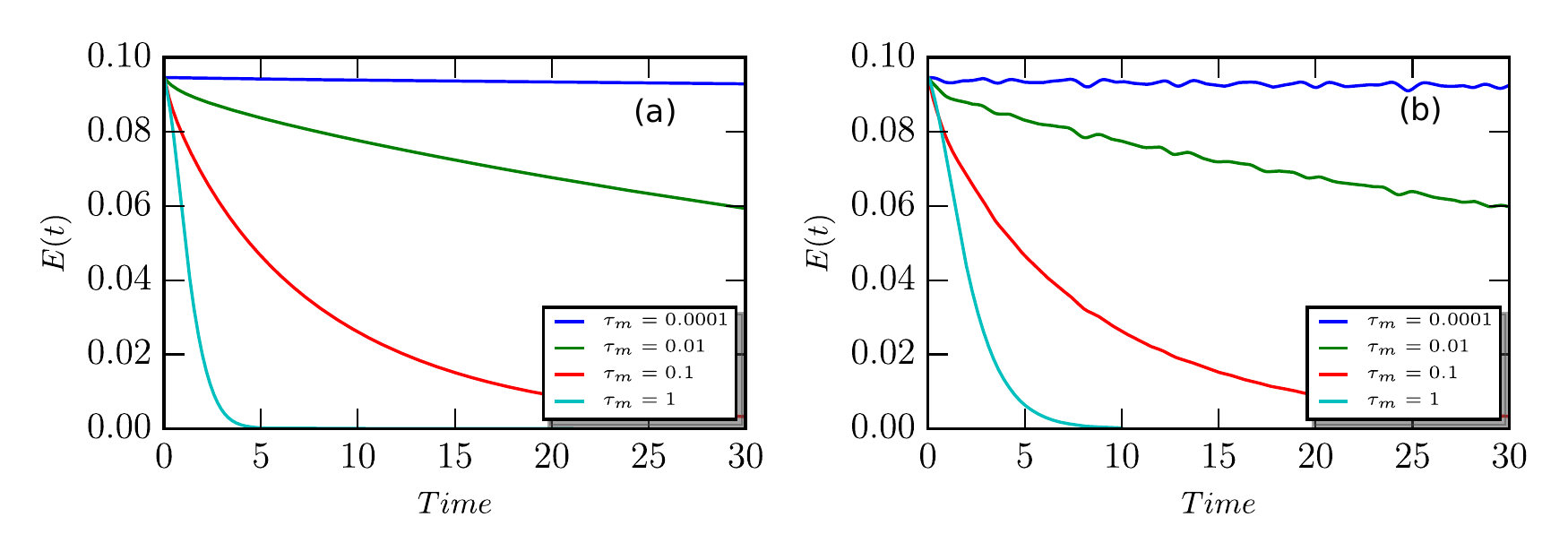}
\caption{(Color online) (a) Time evolution of total kinetic energy of the system for $M = 0.04$ and $\sqrt{\nu/ \tau_m} = 1.412$, (b) Time evolution of total kinetic energy of the system for $M = 0.4$ and $\sqrt{\nu/ \tau_m} = 1.412$.}
\label{kinetic_energy_incomp}
\end{figure*}

The viscous effects acting on the dusty plasma take away the kinetic energy of the system leading to an overall decay trend in the Fig.~\ref{kinetic_energy_incomp}(a). Here the transverse speed $\sqrt{\frac{\nu}{\tau_m}}$ is kept constant ($=\sqrt{2}$) for different values of $\tau_m$. Thus, for higher values of $\tau_m$ the magnitude of viscosity ($\nu$) increases thereby leading to a faster fall in the kinetic energy in Fig.~\ref{kinetic_energy_incomp}(a)\\

\begin{figure*}[h!]
\includegraphics[scale=2.0]{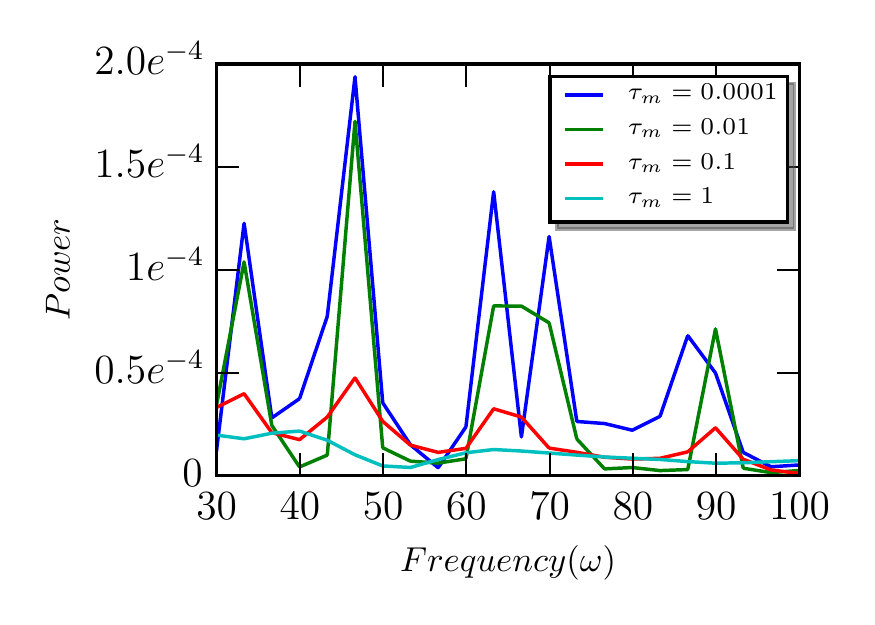}
\caption{(Color online) Frequency of kinetic energy for $M = 0.4$ and $\tau_m = 10^{-4}, 10^{-2}, 10^{-1}, 1$.}
\label{frequency}
\end{figure*}

\begin{figure*}[htbp]
	\begin{center}
	\includegraphics[scale=1]{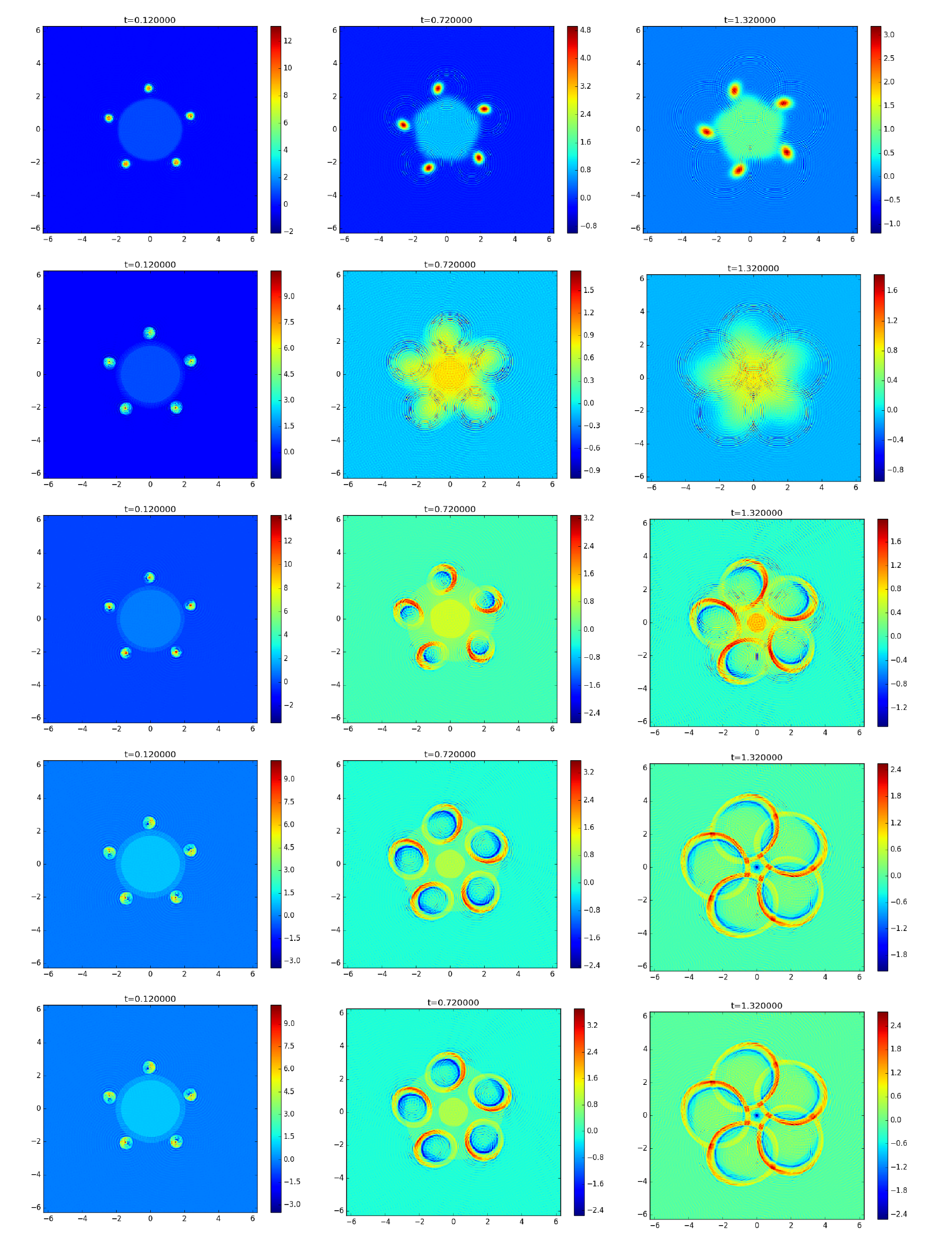}
	
	\caption{(Color online) Time evolution of the prearranged vortex merger with $N_{pv} = 5$, $M = 0.4$ and $\sqrt{\frac{\nu}{\tau_m}} = 1.412$ for grid size $256^2$. The first, second, third, fourth, and fifth rows indicates $\tau_m = 10^{-2}, 10^{-1}, 1, 3, 5$ respectively.}
	\label{snap_tau_m_comp}
	\end{center}
\end{figure*}

The effect of compressibility on such a pre-arranged vortex pattern in a neutral fluid was studied earlier.\cite{rupak:2018} Here, we study the effect of compressibility on the same vortex structure for a viscoelastic dusty plasma. The transverse wave speed ($\frac{\nu}{\tau_m} = 1.412$) has been kept constant. The kinetic energy for different values of $\tau_m$ is shown in Fig.~\ref{kinetic_energy_incomp}(b). The fall in kinetic energy for higher values of $\tau_m$ is identical to that observed for the incompressible case. But we observe some oscillations in the kinetic energy for $\tau_m = 10^{-4}$ and $10^{-2}$. The oscillations are less prominent for $\tau_m = 10^{-1}$ and almost no oscillation can be found for $\tau_m = 1$. We detrend the compressible kinetic energy data from the incompressible one and remove the viscous dissipation for all values of $\tau_m$. Further Fourier transform of the kinetic energy data has been done (using FFTW library\cite{FFTW3:2005}) for different values of $\tau_m$. The Fourier transform of the detrended kinetic energy for $M = 0.4$ is plotted in Fig.~\ref{frequency}. For $\tau_m = 10^{-4}$ and $10^{-2}$ five prominent peaks are observed. The power of the peaks are little less for $\tau_m = 10^{-2}$ than $\tau_m = 10^{-4}$. For $\tau_m \rightarrow 0$, the origin of the peaks can be understood as the interaction of the point-like vortices with the patch vortex. The natural frequency of the sonic wave can be matched with the first peak of the frequency spectrum in Fig.~\ref{frequency}. The other four peaks are the harmonics and the beats with the harmonics. For $\tau_m = 10^{-1}$ a sharp fall in the power of the peaks is observed compared to lower values of $\tau_m$. Finally, for $\tau_m = 1$ we do not observe any noteworthy peak in the kinetic energy spectrum. This helps us to conclude that, for a fixed Mach number ($M = 0.4$) the frequencies generated in the time evolution of total kinetic energy get suppress as the viscoelasticity ($\tau_m$) is increased. In Fig.~\ref{snap_tau_m_comp} the time evolution of the prearranged vortex merger with $N_{pv} = 5$, $M = 0.4$ and $\sqrt{\nu/\tau_m} = 1.412$. The first, second, third, fourth  and fifth rows indicates $\tau_m = 10^{-2}, 10^{-1}, 1, 3, 5$ respectively. To fix the value of $\sqrt{\nu/\tau_m}$ value for each values of $\tau_m$, we change the kinematic  viscosity accordingly. Therefore, for larger values of $\tau_m$, $\nu$ is large and hence, leads to the fast viscous dissipation and vortex mixing.  \\

\section{Summary}
\label{sec-summar}
To summarize, we reported on numerical results obtained from the spatiotemporal evolution of localized coherent vortex flow using generalized hydrodynamic fluid model. In case of single rotating vortex flow (Gaussian vortex), a transverse wave is generated from the localized vortex source and due to viscoelasticity and compressibility of the medium evolution time of generated waves is found to be delayed. A single Rankine vortex patch, which is a solution of Navier-Stokes fluids has also been studied for various values of $\tau_{m}$. In the presence of finite viscoelasticity,  a Rankine vortex gets
de-stabilized by itself and generates several patterns as
the time is evolved. \\

 We have also studied the time evolution of a pre-arranged set of vortices of two different radii of vortices and have observed the merger process. The merger phenomena have been found to be strikingly different from that of a non-viscoelastic fluid. Transverse waves arising out of the point-like vortices around the patch vortex begin to expand and collide with each other giving rise to regions of higher vorticity. The effect of viscoelasticity is found to diminish the frequencies generated in a compressible system. A similar study for different density and interaction potential can be a case of interesting study in future. The central patch like dust vortex itself gets destabilized even without the perturbation from the point-like vortices.
In the limit, $\tau_m \rightarrow 0$ some frills appear out of the point-like vortices. At-large $\tau_m$ the frills get prominent and form wave-front like structures and propagate in time. 


\section{Acknowledgement}
All works have been performed at HPC2013-IITK cluster of IIT Kanpur and Uday and Udbhav clusters at Institute for Plasma Research.  This research was supported in part by the International Centre for Theoretical Sciences (ICTS) during a visit for participating in the program - Turbulence from Angstroms to light years (Code: ICTS/Prog-taly2018/01). RM thanks Samriddhi Sankar Ray at International Center for Theoretical Sciences, India for his initial help regarding pseudo-spectral simulation. RM acknowledges helpful discussions with Prasad Perlekar at TIFR Hyderabad. RM is indebted to Anantanarayanan Thyagaraja at University of Bristol and Govind Krishnaswami at Chennai Mathematical Institute for their several thoughtful suggestions.

\nocite{}

\begin{thebibliography}{10}

\bibitem{morfill:2009}
Gregor~E Morfill, Alexei~V Ivlev, {\it Reviews of modern physics}
  \textbf{2009}, {\it 81} (4), 1353.

\bibitem{tsytovich:2007}
Vadim~N Tsytovich, Gregor Morfill, Sergey~V Vladimirov, Hubertus~M Thomas, {\it
  Elementary physics of complex plasmas}, {\it \bblvol{} 731}, Springer Science
  \& Business Media, \textbf{2007}.

\bibitem{fortov:2009}
Vladimir~E Fortov, Gregor~E Morfill, {\it Complex and dusty plasmas: from
  laboratory to space}, CRC Press, \textbf{2009}.

\bibitem{bonitz:2010}
Michael Bonitz, Norman Horing, Patrick Ludwig, {\it Introduction to complex
  plasmas}, {\it \bblvol{}~59}, Springer Science \& Business Media,
  \textbf{2010}.

\bibitem{shukla:2015}
Padma~K Shukla, A~A Mamun, {\it Introduction to dusty plasma physics}, CRC
  Press, \textbf{2015}.

\bibitem{morfillRMP}
Gregor~E. Morfill, Alexei~V. Ivlev, {\it Rev. Mod. Phys.} \textbf{2009}, {\it
  81}, 1353.

\bibitem{fusion}
U.~de~Angelis, {\it Physics of Plasmas} \textbf{2006}, {\it 13} (1), 012514.

\bibitem{rao}
N.~N Rao, P.~K Shukla, M.~Y Yu, {\it Planet. Space. Sci} \textbf{1990}, {\it
  38}, 543.

\bibitem{pustejovska:2008}
P~Pustejovsk{\'a}, {\it WDS'08 Proceedings of Contributed Papers, Part}
  \textbf{2008}, {\it 3}, 32--37.

\bibitem{kaw:1998}
P~K Kaw, A~Sen, {\it Physics of Plasmas} \textbf{1998}, {\it 5} (10),
  3552--3559.

\bibitem{diaw:2015}
Abdourahmane Diaw, Michael~Sean Murillo, {\it Physical Review E} \textbf{2015},
  {\it 92} (1), 013107.

\bibitem{luo:2016}
Di~Luo, Bin Zhao, GuangYue Hu, Tao Gong, YuQing Xia, Jian Zheng, {\it Physics
  of Plasmas} \textbf{2016}, {\it 23} (5), 052707.

\bibitem{banerjee:2012}
D~Banerjee, M~S Janaki, N~Chakrabarti, {\it Physical Review E} \textbf{2012},
  {\it 85} (6), 066408.

\bibitem{sanat:2012}
Sanat~Kumar Tiwari, Amita Das, Dilip Angom, Bhavesh~G Patel, Predhiman Kaw,
  {\it Physics of Plasmas} \textbf{2012}, {\it 19} (7), 073703.

\bibitem{akanksha:2014}
Akanksha Gupta, R~Ganesh, Ashwin Joy, {\it Physics of Plasmas} \textbf{2014},
  {\it 21} (7), 073707.

\bibitem{akanksha:2017}
Akanksha Gupta, \bblphdthesis{}, Institute for Plasma Research, Homi Bhabha
  National Institute, \textbf{2017}.

\bibitem{akanksha:2018}
Akanksha Gupta, Rajaraman Ganesh, {\it Physics of Plasmas} \textbf{2018}, {\it
  25} (1), 013705.

\bibitem{sandeep:2017}
Sandeep Kumar, Bhavesh Patel, Amita Das, {\it arXiv preprint arXiv:1710.04001}
  \textbf{2017}.

\bibitem{saitou:2012}
Yoshifumi Saitou, Yoshiharu Nakamura, Tetsuo Kamimura, Osamu Ishihara, {\it
  Physical review letters} \textbf{2012}, {\it 108} (6), 065004.

\bibitem{nakamura:1999}
Y~Nakamura, H~Bailung, P~K Shukla, {\it Physical review letters} \textbf{1999},
  {\it 83} (8), 1602.

\bibitem{meyer:2013}
John~K Meyer, Robert~L Merlino, {\it Physics of Plasmas} \textbf{2013}, {\it
  20} (7), 074501.

\bibitem{surabhi:2016}
Surabhi Jaiswal, P~Bandyopadhyay, A~Sen, {\it Physical Review E} \textbf{2016},
  {\it 93} (4), 041201.

\bibitem{shukla:2003}
P~K Shukla, A~A Mamun, {\it New Journal of Physics} \textbf{2003}, {\it 5} (1),
  17.

\bibitem{hasegawa:2004}
Akira Hasegawa, K~P Shukla, {\it Physics Letters A} \textbf{2004}, {\it 332}
  (1-2), 82--85.

\bibitem{vranjevs:1999}
J~Vranje{\v{s}}, G~Mari{\'c}, K~P Shukla, {\it Physics Letters A}
  \textbf{1999}, {\it 258} (4-6), 317--322.

\bibitem{wilms:2017}
Jochen Wilms, Alexander Piel, {\it Physics of Plasmas} \textbf{2017}, {\it 24}
  (8), 083703.

\bibitem{krommes:2002}
John~A Krommes, {\it Physics Reports} \textbf{2002}, {\it 360} (1-4), 1--352.

\bibitem{frisch:1995}
Uriel Frisch, {\it Turbulence: the legacy of AN Kolmogorov}, Cambridge
  university press, \textbf{1995}.

\bibitem{rupak:2018}
Rupak Mukherjee, Akanksha Gupta, Rajaraman Ganesh, {\it arXiv preprint
  arXiv:1802.03240} \textbf{2018}.
  
\bibitem{rupak:2018:IEEE}
Rupak Mukherjee, Rajaraman Ganesh, Vinod Saini, Udaya Maurya, Nagavijayalakshmi Vydyanathan, Bharatkumar Sharma, {\it arXiv preprint
  arXiv:1810.12707} \textbf{2018}.

\bibitem{fine:1995}
K. S. Fine, A. C. Cass, W. G. Flynn, C. F. Driscoll, {\it
  Physical Review Letters} \textbf{1995}, {\it 75} (18), 3277.

\bibitem{durkin:2000}
D~Durkin, J~Fajans, {\it Physical review letters} \textbf{2000}, {\it 85} (19),
  4052.

\bibitem{driscoll:1999}
C.F. Driscoll, D.A. Schecter, D.Z. Jin, D.H.E. Dubin, K.S. Fine, A.C. Cass, {\it Physica A} \textbf{1999}, {\it 263}, 284-292

\bibitem{schecter:1999}
D. A. Schecter, D. H. E. Dubin, K. S. Fine, C. F. Driscoll,  {\it Physics of Fluids} \textbf{1999}, {\it 11} (4), 905

\bibitem{ganesh:2002}
R~Ganesh, J~K Lee, {\it Physics of Plasmas} \textbf{2002}, {\it 9} (11),
  4551--4559.

\bibitem{perlekar:2016}
Rohith~V Swaminathan, S~Ravichandran, Prasad Perlekar, Rama Govindarajan, {\it
  Physical Review E} \textbf{2016}, {\it 94} (1), 013105.

\bibitem{kaw}
P.~K. Kaw, A.~Sen, {\it Physics of Plasmas} \textbf{1998}, {\it 5}, (10), 3552.

\bibitem{vlad}
Vladimir~E. Fortov, Vladimir~I. Molotkov, Anatoli~P. Nefedov, Oleg~F. Petrov,
  {\it Physics of Plasmas} \textbf{1999}, {\it 6} (5), 1759--1768.

\bibitem{FFTW3:2005}
Matteo Frigo, Steven~G. Johnson, {\it Proceedings of the IEEE} \textbf{2005},
  {\it 93} (2), 216--231, Special issue on ``Program Generation, Optimization,
  and Platform Adaptation''.

\bibitem{iov}
Michele Iovieno, Carlo Cavazzoni, Daniela Tordella, {\it Computer Physics
  Communications} \textbf{2001}, {\it 141} (3), 365 -- 374.

\bibitem{orszag}
G.~S. Patterson, Steven~A. Orszag, {\it Physics of Fluids (1958-1988)}
  \textbf{1971}, {\it 14} (11), 2538--2541.

\bibitem{ea}
E.~A. Coutsias, F.~R. Hansen, T.~Huld, G.~Knorr, J.~P. Lynov, {\it Physica
  Scripta} \textbf{1989}, {\it 40} (3), 270.

\bibitem{terakado:2014}
Daiki Terakado, Yuji Hattori, {\it Physics of Fluids} \textbf{2014}, {\it 26}
  (8), 085105.

\bibitem{bayly:1992}
B~J Bayly, CD~Levermore, T~Passot, {\it Physics of Fluids A: Fluid Dynamics}
  \textbf{1992}, {\it 4} (5), 945--954.

\bibitem{vikram:2014}
Vikram Singh~Dharodi, Sanat Kumar~Tiwari, Amita Das, {\it Physics of Plasmas}
  \textbf{2014}, {\it 21} (7), 073705.

\bibitem{deem:1978}
Gary~S Deem, Norman~J Zabusky, {\it Physical Review Letters} \textbf{1978},
  {\it 40} (13), 859.

\bibitem{lansky:1997}
M~Lansky, I, T~M O'Neil, D~A Schecter, {\it Physical review letters}
  \textbf{1997}, {\it 79} (8), 1479.

\end{thebibliography}
\providecommand{\noopsort}[1]{}\providecommand{\singleletter}[1]{#1}%
\providecommand{\url}[1]{\texttt{#1}}
\providecommand{\urlprefix}{}
\providecommand{\foreignlanguage}[2]{#2}
\providecommand{\Capitalize}[1]{\uppercase{#1}}
\providecommand{\capitalize}[1]{\expandafter\Capitalize#1}
\providecommand{\bibliographycite}[1]{\cite{#1}}
\providecommand{\bbland}{and}
\providecommand{\bblchap}{chap.}
\providecommand{\bblchapter}{chapter}
\providecommand{\bbletal}{et~al.}
\providecommand{\bbleditors}{editors}
\providecommand{\bbleds}{eds: }
\providecommand{\bbleditor}{editor}
\providecommand{\bbled}{ed.}
\providecommand{\bbledition}{edition}
\providecommand{\bbledn}{ed.}
\providecommand{\bbleidp}{page}
\providecommand{\bbleidpp}{pages}
\providecommand{\bblerratum}{erratum}
\providecommand{\bblin}{in}
\providecommand{\bblmthesis}{Master's thesis}
\providecommand{\bblno}{no.}
\providecommand{\bblnumber}{number}
\providecommand{\bblof}{of}
\providecommand{\bblpage}{page}
\providecommand{\bblpages}{pages}
\providecommand{\bblp}{p}
\providecommand{\bblphdthesis}{Ph.D. thesis}
\providecommand{\bblpp}{pp}
\providecommand{\bbltechrep}{}
\providecommand{\bbltechreport}{Technical Report}
\providecommand{\bblvolume}{volume}
\providecommand{\bblvol}{Vol.}
\providecommand{\bbljan}{January}
\providecommand{\bblfeb}{February}
\providecommand{\bblmar}{March}
\providecommand{\bblapr}{April}
\providecommand{\bblmay}{May}
\providecommand{\bbljun}{June}
\providecommand{\bbljul}{July}
\providecommand{\bblaug}{August}
\providecommand{\bblsep}{September}
\providecommand{\bbloct}{October}
\providecommand{\bblnov}{November}
\providecommand{\bbldec}{December}
\providecommand{\bblfirst}{First}
\providecommand{\bblfirsto}{1st}
\providecommand{\bblsecond}{Second}
\providecommand{\bblsecondo}{2nd}
\providecommand{\bblthird}{Third}
\providecommand{\bblthirdo}{3rd}
\providecommand{\bblfourth}{Fourth}
\providecommand{\bblfourtho}{4th}
\providecommand{\bblfifth}{Fifth}
\providecommand{\bblfiftho}{5th}
\providecommand{\bblst}{st}
\providecommand{\bblnd}{nd}
\providecommand{\bblrd}{rd}
\providecommand{\bblth}{th}

\end{document}